\documentclass[12pt]{article}
\usepackage[margin=0.95 in]{geometry}
\usepackage{bm}
\usepackage{amsthm}
\usepackage{breqn}
\usepackage{amssymb,amsfonts}
\usepackage[all]{xy}
\usepackage{graphicx}
\usepackage{amsmath}
\usepackage{amssymb}
\usepackage{float}
\usepackage{array}
\usepackage{mathtools}
\usepackage{mathrsfs} 
\usepackage{comment}
\usepackage[hyperfootnotes=false]{hyperref}
\usepackage{cite}
\numberwithin{equation}{section}
\usepackage{xfrac}
\usepackage{caption}
\usepackage{slashed}
\usepackage{dsfont}
\usepackage[all]{xy}
\usepackage{graphicx}
\numberwithin{equation}{section}
\numberwithin{equation}{section}
\numberwithin{table}{section}

\newcommand{\curlyA}{\mathcal{A}}
\newcommand{\curlyB}{\mathcal{B}}
\newcommand{\curlyC}{\mathcal{C}}

\newcommand{\curlyN}{\mathcal{N}}

\newcommand{\PP}{\mathbb{P}}

\begin{document}

\hypersetup{pageanchor=false}
\begin{titlepage}
\vbox{\halign{#\hfil    \cr}}  
\vspace*{15mm}
\begin{center}
{\Large \bf Symmetric Group Gauge Theories }

\vspace*{3mm}

{\Large \bf and  Simple  Gauge/String Dualities}

\vspace*{10mm} 

{\renewcommand*{\thefootnote}{\fnsymbol{footnote}}
{\large Lior Benizri and Jan Troost}
\vspace*{8mm}
}
\setcounter{footnote}{0}

Laboratoire de Physique de l'\'Ecole Normale Sup\'erieure \\ 
\hskip -.05cm
CNRS, ENS, Universit\'e PSL,  Sorbonne Universit\'e, \\
Universit\'e  Paris Cit\'e F-75005 Paris, France\\
\vspace{6mm}
%

\vspace*{8mm}

\end{center}

\begin{abstract} {We study two-dimensional topological gauge theories with gauge group equal to the symmetric group $S_n$ and their string theory duals. The simplest such theory is the topological quantum field theory of principal $S_n$ fiber bundles. Its correlators are equal to Hurwitz numbers. The operator products in the gauge theory for each finite value of $n$ are coded in one partial permutation algebra. 
We propose a generalization of the partial permutation algebra to the symmetric orbifold topological quantum field theory 
of any seed theory and show that the theory factorizes into marked partial permutation combinatorics and seed Frobenius algebra properties.
Moreover, we exploit the established correspondence between Hurwitz theory and the stationary sector of Gromov-Witten theory on the sphere  to prove an exact gauge/string duality. The relevant field theory is a grand canonical version of Hurwitz theory and its two-point functions are obtained by summing over all values of the instanton degree of the maps covering the sphere. We stress that one must look for a multiplicative basis on the boundary to match the bulk operator algebra of single string insertions. The relevant boundary observables are completed cycles. 
}  
\end{abstract}

\end{titlepage}

\hypersetup{pageanchor=true}

\setcounter{tocdepth}{2}
\tableofcontents

\section{Introduction}
The large $N$ limit of four-dimensional gauge theories with $U(N)$ gauge group permits an organization of the Feynman diagram expansion in terms of Riemann surfaces of increasing genus \cite{tHooft:1973alw}. This feeds the hope that there is a string theory dual to confining and other four-dimensional gauge theories which efficiently describes the physics at large or even moderate values of $N$. One concrete realization of this idea is the holographic duality between ${\cal N}=4$ super Yang-Mills theory with $U(N)$ gauge group and string theory in $AdS_5 \times S^5$ with $N$ units of flux \cite{Maldacena:1997re}.
However, it remains hard to prove the duality even for this conformal theory from first principles, in particular beyond the planar limit.
In two dimensions, 't Hooft's ideas are also applicable. There is a genus expansion of the observables of $U(N)$ gauge theories in two dimensions \cite{tHooft:1973alw,Gross:1992tu,Gross:1993yt,Gross:1993hu}.\footnote{See e.g. \cite{Aharony:2023tam} for a recent contribution to the  subject.} The string theoretic description in terms of integrals over the moduli spaces of Riemann surfaces \cite{Cordes:1994fc} is intricate but  breeds hope for a simple gauge/string duality.  
At least in two dimensions, there is a further suggestion that the genus expansion  also applies to gauge theories with a discrete gauge group equal to the symmetric group $S_n$. This can be argued on the basis of Higgsing the $U(N)$ symmetry, through Schur-Weyl duality or through holographic dualities in  string theory  that suggest an equivalence between  $AdS_3$ string theory and a symmetric orbifold conformal field theory \cite{Maldacena:1997re,El-Showk:2011yvt}. 
It is natural to ask  whether at least one such gauge/string duality can be established exactly.  One plan of this paper is to explain that there is \cite{OP1}. For other  valuable contributions in this direction see  \cite{Jevicki:1998bm,Gopakumar:2011ev,Li:2020zwo,Eberhardt:2019ywk,Eberhardt:2021vsx} among others as well as references therein.

Before we delve into the bulk of the paper, we propose an overview of the central concepts and results and where to find them in the paper. Firstly, we discuss the simplest gauge theory with symmetric gauge group $S_n$ in section \ref{HurwitzTheory}. It is a theory of principal $S_n$ bundles on Riemann surfaces which counts $n$-fold branched coverings.  It is known that the combinatorics of these coverings is efficiently captured by an algebra called the partial permutation algebra \cite{IvanovKerov}. A partial permutation is a pair consisting of a subset of $\{ 1, 2, \dots, n \}$ along with a permutation that permutes a further subset of its elements. An example partial permutation for $n=4$ is $(\{ 1,2,3 \},(12))$. Furthermore, the limit for $n \rightarrow \infty$ of the partial permutation algebra exists and it encodes all of the information on the operator algebra of the gauge theory for every finite value of $n$ \cite{IvanovKerov}. In fact, it contains more information since the combinatorial properties of partial permutations are more refined than those of permutations. 
At the end of section \ref{HurwitzTheory}, we exhibit
a multiplicative basis of the algebra 
which is important to the further developments in the paper and in particular the string theory interpretation of the gauge theory in section \ref{H/GW}.

The Hurwitz theory can be thought of as the topological symmetric orbifold theory of a trivial seed theory. In section  \ref{SecondQuantizedtopological quantum field theory} we generalize our observations on the Hurwitz theory to generic two-dimensional topological symmetric orbifold theories.
We stress that the embedding of the finite $n$ theories in the infinite $n$ algebras retains all information and encodes it more efficiently.
To that end, we introduce and analyze novel algebras of observables. We are able to largely factorize the symmetric group combinatorics and the seed Frobenius algebra calculations. In the process we identify a hierarchy of combinatorics which refines the partial permutation algebra combinatorics even further. We thus have a refinement of Hurwitz combinatorics to partial permutation combinatorics and furthermore to the combinatorics of partial permutations marked by operators in a seed Frobenius algebra. Section \ref{SecondQuantizedtopological quantum field theory} also contains concrete correlator calculations to illustrate these concepts as well as the sketch of a Feynman diagram calculus that simplifies the calculation of generic topological symmetric orbifold correlators. 

The Gromov-Witten/Hurwitz correspondence \cite{OP1} is the subject of section \ref{H/GW}. We explain how the grand canonical perspective on the Hurwitz theory, in which we sum over all values of the degree of the cover $n$ is essential in obtaining a dual string theory interpretation. Indeed, the introduction of the grand canonical viewpoint and the infinite $n$ permutation algebra allow for the interpretation of the theory in terms of a sum over string world sheet instantons. Combining this with the existence of a multiplicative basis with the right symmetry properties allows us to interpret the correspondence as a gauge/string duality. 
The detailed mathematical correspondence realizes physical intuitions and modifies them in important details. In particular, we stress that the duals of bulk single string vertex operators are completed cycles \cite{OP1}, a particular linear combination of partial permutations. Moreover, we argue that the instanton sum in the Gromov-Witten theory on $\mathbb{P}^1$ gives rise to a non-degenerate metric on a basis of partial permutations. We briefly speculate on how the duality could generalize to the much broader setting of topological symmetric orbifold theories. We conclude the paper in section \ref{Conclusions} with a summary and a discussion of open problems. Our paper is peppered with remarks that may make the largely mathematical literature on the subject more palatable.

\section{\texorpdfstring{Symmetric Group Gauge Theories}{}}
\label{HurwitzTheory}
\label{Hurwitz}
In this section, we briefly review two-dimensional topological 
field theories and their equivalence to Frobenius algebras. See e.g. \cite{DijkgraafPhD,Atiyah:1989vu,Moore,DVV,Kock} for excellent introductions 
and summaries. We then concentrate on the Dijkgraaf-Witten theory \cite{Dijkgraaf:1989pz} of principal $S_n$ bundles on Riemann surfaces. We show that the algebra of partial permutations \cite{IvanovKerov} encodes all information on this theory for every value of $n$. We moreover introduce and motivate a multiplicative basis of the partial permutation algebra that will play an important role in section \ref{GaugeStringDuality}. 
Subsections \ref{TQFTFA} to \ref{topological quantum field theory} are an efficient review of standard facts spread over the mathematics and physics literature. Subsection \ref{PartialPermutationsAndExtendedAlgebra} has more originality since it makes considerably more explicit how the infinite $n$
partial permutation algebra of \cite{IvanovKerov} plays a crucial role in interpreting the grand canonical Hurwitz theory as a topological quantum field theory, a point left largely implicit in \cite{OP1}.

\subsection{Topological Quantum Field Theories and Frobenius Algebras}
\label{TQFTFA}
\begin{figure}[ht]
	\centering
	\includegraphics[width=2cm]{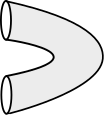}
	\caption{The two-point function defines a metric.}
 \label{TwoPointFunction}
\end{figure}

\begin{figure}[ht]
	\centering
	\includegraphics[width=6cm]{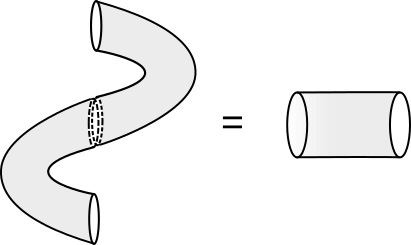}
	\caption{The snake diagram is equivalent to the identity morphism.}
 \label{Snake}
\end{figure}

\begin{figure}[ht]
	\centering
	\includegraphics[width=10cm]{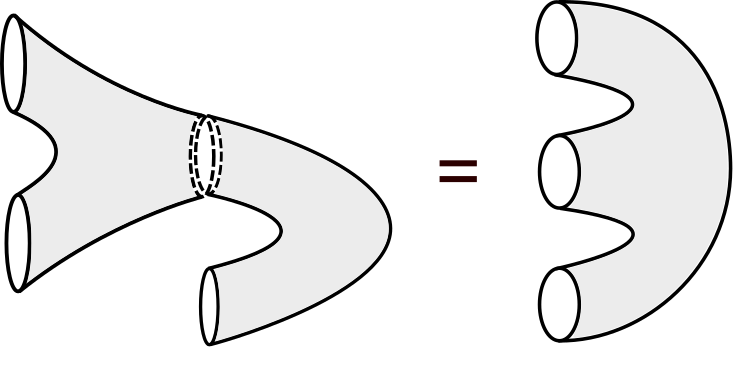}
 \caption{The composition of the product and metric yields a three-point function.}
 \label{ProductAndThreePointFunction}
\end{figure}

Topological field theories are quantum field theories for 
which the observables are independent of the metric on the manifold.
A $d+1$-dimensional (closed) topological quantum field theory is defined 
from the mathematical perspective \cite{Atiyah:1989vu} as a functor $Z$ 
between the category of cobordisms {Cob}$(d)$ and the category of vector spaces {Vect}.
The objects in the category {Cob}$(d)$ are closed $d$-dimensional manifolds. 
A morphism $\phi:M\to N$ is a smooth oriented $d+1$-dimensional manifold whose
boundaries are given by $M$ and $N$, which are manifolds of opposite orientation.
The objects of the category {Vect} are  vector spaces and its morphisms are linear maps. The vector spaces play the role of the spaces of ingoing and outgoing states in a quantum field theory. When the total manifold is closed, the functor computes a partition function. A  consequence of the gluing properties of manifolds is that the correlation functions of the topological theory factorize. Indeed, consider the two-point function, which is represented in Figure \ref{TwoPointFunction}.
One can show that the bilinear form $\eta$ is non-degenerate by composing with the reversed cobordism $\eta^{-1}:\mathbb{C}\to V\times V$ \cite{Moore} and arguing that the resulting snake diagram is equivalent to the identity morphism or cylinder, see figure \ref{Snake}.
The non-degeneracy guarantees that the form $\eta$ can serve as a metric. All correlation functions $\langle \phi_1 \phi_2 ... \phi_n \rangle_g$ on a genus $g$ surface factorize \cite{DVV}:
\begin{equation}
	\langle \phi_1 \phi_2 ... \phi_n \rangle_g=\sum_{i,j} \langle \phi_1 ... \phi_k \phi_i  \rangle_{g_1} \eta^{ij} \langle \phi_j \phi_{k+1} ... \phi_n \rangle_{g_2} \, .
	\label{RecRel1}
\end{equation}
The sum is over all the states in the theory and $g_1, g_2$ are two arbitrary
genera that sum up to the total genus $g$. 
Moreover, we can also lower the genus of a diagram by cutting a loop:
\begin{equation}
	\langle \phi_1 \phi_2 ... \phi_n \rangle_g=\sum_{i,j} \eta^{ij} \langle \phi_i \phi_j \phi_1 ... \phi_n \rangle_{g-1} \, .
	\label{RecRel2}
\end{equation}
These degeneration formulas reduce the computation of $n$-point functions on a genus
$g$ surface to that of two- and three-point functions at genus zero. It is therefore sufficient to study the latter to fully characterize the  theory. 
The operators of a closed topological field theory exhibit the structure of a commutative Frobenius algebra $A$ \cite{DijkgraafPhD}. 
We explicitly parameterize the operator product 
\begin{equation}
	\phi_i  \phi_j=c_{ij}^{\ \; k} \phi_k \, 
\end{equation}
in terms of the structure constants  $c_{ij}^{\ k}$ of the  algebra $A$. They are closely related to the three-point functions, through composition with the metric:
\begin{equation}
	\langle \phi_i \phi_j \phi_k \rangle_0 = c_{ijk}=c_{ij}^{\ \; l}\eta_{lk} \, .
\end{equation}
See Figure \ref{ProductAndThreePointFunction}.
We stress that specifying a topological field theory in two dimensions is equivalent to specifying its structure constants $c_{ij}^{\ \ k}$ and its metric $\eta_{ij}$. The former defines the operator algebra structure on the theory and the latter is the basic ingredient required to compute correlation functions from the operator algebra.
It is useful  to recall the full algebraic structure of a Frobenius algebra. A Frobenius algebra $A$ is an associative, unital algebra equipped with a non-degenerate bilinear form 
$\eta$ that satisfies:
\begin{equation}
	\eta(ab,c)=\eta(a,bc) \, 
	\label{Frobenius}
\end{equation}
for all elements $a,b$ and $c$ in the algebra $A$. 
The algebra reduces $n$-point functions to one-point functions and the one-point functions are evaluated using the topological metric $\eta$
\begin{equation}
	\langle \phi \rangle = \langle\phi \,  1_{A} \rangle = \eta(\phi,1_A)\ ,
\end{equation}
where $1_A$ denotes the unit in the algebra $A$. 
This motivates an alternative and equivalent definition of a commutative Frobenius algebra: it 
is a unital, associative algebra equipped with a trace $T$ whose quadratic form 
gives rise to a metric $\eta(a,b)=T(a b)$.
In order to incorporate fermions, a slight generalization is required. The
appropriate algebraic structure is that of a $\mathbb{Z}_2$ graded
commutative algebra.\footnote{We will not track the associated signs in this paper.}

\subsection{Gauge Theories with a Finite Gauge Group}
We concentrate on gauge theories with a finite gauge group or two-dimensional Dijkgraaf-Witten theories 
\cite{Dijkgraaf:1989pz}.\footnote{See also \cite{Gardiner:2020vjp,Blau:1991mp} for pedagogical introductions.} In these theories, the path integral is a sum over equivalence classes of principal $G$-bundles on the connected spacetime manifold $M$. Let $\mathcal{C}_M$ denote the set of all inequivalent principal $G$-bundles. The set $\mathcal{C}_M$ is in bijection with the set of gauge equivalence classes of morphisms from the fundamental group $\pi_1(M)$ of the manifold $M$ to the group $G$. The morphism $\phi:\pi_1(M)\to G$ and its conjugate $g\phi g^{-1}$ correspond to equivalent bundles. We thus have
\begin{equation}
	\mathcal{C}_M \equiv Hom(\pi_1(M),G)/G\, .
\end{equation}
We denote by $p$ the projection induced by the quotient on the right-hand side. There is a natural measure $\mu$
on this orbifold space that is given by
\begin{equation}
	\mu([\phi]):=\frac{\lvert p^{-1}([\phi]) \rvert}{\lvert G \rvert}=\frac{1}{\lvert \text{Stab}(\phi) \rvert} \, ,
\end{equation}
where $\text{Stab}(\phi)$ is the subgroup of $G$ that leaves $\phi$ invariant. 
The partition function $Z$ on the manifold $M$ thus reads
\begin{equation}
	Z[M]=\sum_{[\phi] \in C_M}\frac{1}{\lvert \text{Stab}(\phi) \rvert}=\frac{\lvert Hom(\pi_1(M),G)\rvert }{\lvert G\rvert } \, .
\end{equation}
This last expression simplifies using Mednykh's formula \cite{MedFormula,MednykhsProof}:
\begin{equation}
	\lvert Hom(\pi_1(M_g),G)\rvert=\lvert G \rvert \sum_{\lambda} \left(\frac{d_{\lambda}}{\lvert G \rvert}\right)^{2-2g} \, ,
\end{equation}
where the sum is over all irreducible representations $\lambda$ of the group $G$ and $g$ is the genus of the surface $M=M_g$. One
finds the partition function:\footnote{We note the similarity to the partition functions of two-dimensional Yang-Mills theories. }
\begin{equation}
Z[M_g] = \sum_{\lambda} \left(\frac{d_{\lambda}}{\lvert G \rvert}\right)^{2-2g} \, .
\end{equation}
The reasoning and formula can be generalized to compute all correlation 
functions in a Dijkgraaf-Witten theory \cite{Dijkgraaf:1989pz,Gardiner:2020vjp,Blau:1991mp}. Operator
insertions can be thought of as half cylinders extending from the manifold. They can be conformally mapped to an operator insertion at a point on the Riemann surface. The gauge invariant operators are determined by the conjugacy classes of the holonomy of the gauge field around the half cylinder. The partition function for a Riemann surface $M_{g}$
of genus $g$ with $k$ insertions is:
\begin{equation}
Z[M_{g},\xi_i]=\sum_{\lambda}\left(\frac{d_{\lambda}}{\lvert G \rvert}\right)^{2-2g-k}\prod_{i=1}^k \frac{\lvert \xi_i \rvert}{\lvert G \rvert} \chi_{\lambda}(\xi_i)\, ,	
\label{PartitionFunctions}
\end{equation}
where $\xi_i$ are the monodromy conjugacy classes around the $k$ insertions, $d_{\lambda}$ are the dimensions of the irreducible representations $\lambda$ and $|G|$ is the cardinal number of the group while $\chi_{\lambda}$ are the characters of the irreducible representations $\lambda$. The normalized correlation functions of the corresponding operators $\xi_i$ are
\begin{equation}
    \langle \xi_1 \dots \xi_k \rangle_g =\frac{Z[M_{g},\xi_i]}{Z[M_{g}]} \, .
\end{equation}
In this paper, we study these observables.\footnote{There are also open-closed and defect topological quantum field theories with more general observables worth exploring. See e.g. \cite{MooreSegal,Carqueville:2016nqk}.}

\subsection{The Symmetric Group Gauge Theory}
\label{topological quantum field theory}
In this subsection we specify the gauge group $G$ to be equal to the  symmetric group $G=S_n$. We recall the interpretation of the bundles as covers of the manifold and the connection of the correlators to Hurwitz's counting of covers of Riemann surfaces. We then introduce an algebra of operators that is independent of the number $n$ yet captures the properties of the finite $n$ theories with gauge group $S_n$. 
The relevant algebra is the algebra of partial permutations \cite{IvanovKerov}. We review the link of the algebra to the shift symmetric polynomials \cite{IvanovKerov}
and introduce a basis in the latter space that prepares for the identification of a gauge/string duality \cite{OP1} in section \ref{GaugeStringDuality}. 

\subsubsection{The Hurwitz Theory}
The gauge theory correlation functions (\ref{PartitionFunctions}) for the symmetric gauge group $G=S_n$ are equal to Hurwitz numbers. One way to realize this is to note that the counting of homomorphisms from the first fundamental group of the surface into the group $G$ modulo gauge equivalences is equivalent to the Hurwitz enumeration problem of branched covers \cite{CavalieriMiles}. Another way to directly confirm the correspondence is by using the relation between characters and sums of elements in conjugacy classes, and then the relation of the algebra of conjugacy classes to the Hurwitz numbers. In both cases, one winds up with the equation \cite{CavalieriMiles}:
\begin{equation}
Z[M_{g}, \xi_i] = H_{g,n}^{\bullet} (\xi_1,\dots, \xi_k) \, ,
\end{equation}
where the right-hand side are the Hurwitz numbers counting degree $n$ (possibly disconnected) covers of a target closed two-dimensional surface of genus $g$ with $k$ insertions of prescribed ramification type $\xi_i$. The upper index ${\bullet}$  indicates that we allow for disconnected surfaces.
One recognizes in equation (\ref{PartitionFunctions}) the Burnside formula for the Hurwitz numbers  \cite{CavalieriMiles}. The operator insertions of the topological theory correspond to branch points in Hurwitz covers. The monodromy class can thus be viewed as a partition of 
the degree of the cover, with each integer appearing in the partition corresponding to the length of a cycle. In summary, the observables of the 
two-dimensional Dijkgraaf-Witten theory with gauge group $S_n$ may be identified with the conjugacy classes of $S_n$ and their correlation 
functions compute Hurwitz numbers -- this is well-known
\cite{DijkgraafConference,Gunningham:2012qy,Cavalieri:2005sb}.
We will refer to the theory at hand  as the $S_n$ gauge theory or
as the Hurwitz field theory.

\subsubsection{Remarks on Degenerations}
To further explain how Hurwitz theory and the $S_n$ topological quantum field theory match, we rederive some recursion relations for the Hurwitz numbers \cite{CavalieriMiles} and generalize them to surfaces of arbitrary genus
using the property (\ref{RecRel1}) \cite{DVV} :
\begin{align}
	H^{\bullet}_{g,n}(\lambda_1,...,\lambda_k) &=\sum_{i,j} H^{\bullet}_{g_1,n}(\lambda_{1},...,\lambda_s, \lambda_i)\, \eta^{ij} H^{\bullet}_{g_2,n}(\lambda_j,\lambda_{s+1},...,\lambda_k)\\
	&=\sum_{\mu \vdash n} \lvert \text{Stab}(\mu) \rvert H^{\bullet}_{g_1,n}(\lambda_{1},...,\lambda_s, \mu)\,  H^{\bullet}_{g_2,n}(\mu,\lambda_{s+1},...,\lambda_k)\, , 
\end{align}
where $n=\lvert \lambda_i \rvert$ is the degree of the map and $\text{Stab}(\mu)$ is the set of elements that leave a permutation in the conjugacy class corresponding to the partition $\mu$ invariant under conjugation. To go from the first to the second line, we used the expression of the metric on the sphere that we compute explicitly in the next subsection (\ref{MetricOnTheSphere}). 
This formula allows us to reduce the number of branch points in the correlator. Moreover, we can also reduce the genus of the Hurwitz number by applying the cutting property (\ref{RecRel2}) to the Hurwitz theory:
\begin{equation}
	H^{\bullet}_{g,n}(\lambda_1,...,\lambda_k)=\sum_{\mu \vdash n} \lvert \text{Stab}(\mu)\rvert  H^{\bullet}_{g-1,n}(\mu,\mu,\lambda_1,...,\lambda_k) \, .
\end{equation}
\subsubsection{The Hurwitz theory on a sphere}
In the remainder of the paper, we will study the Hurwitz theory on a sphere. Its operator algebra is the algebra $\mathcal{C}_n$ of conjugacy class sums inside the group algebra $\mathbb{C}[S_n]$ of the symmetric group:
\begin{equation}
	\mathcal{C}_n:=\mathbb{C}[S_n]^{S_n}
 \,. 
\end{equation}
The upper $S_n$ index restricts to the invariants of the group algebra $\mathbb{C}[S_n]$ under the action by conjugation. The conjugacy class algebra 
$\curlyC_n$ is identified with the center of the group ring $\mathbb{C}[S_n]$
and is therefore commutative. We refer to this as the Hurwitz (operator) algebra.
The algebra $\mathbb{C}[S_n]$ is semisimple and non-commutative and can be thought of as the algebra of boundary operators \cite{MooreSegal,Fukuma:1993hy}. 
We will denote the operator associated to the conjugacy class $[\pi]$ of the permutation $\pi$ by $C_{\pi}=\sum_{\sigma \in [\pi] \subset S_n} \hspace{-1mm} \sigma$. The set of such operators forms a basis of the Hurwitz algebra $\curlyC_n$. Furthermore, the canonical Frobenius one-form on the group algebra $\mathbb{C}[S_n]$ induces a symmetric, non-degenerate one-form on the Hurwitz algebra $\curlyC_n$ that extracts the coefficient $[C_e]$ of the argument 
\begin{equation}
    \langle C_{\xi} \rangle_{0,n} =\delta_{[\xi],[e]}.
\end{equation}
Once more, the correlation functions of the class operators $C_{\xi_i}$ obtained using this one-point function are the Hurwitz numbers
\cite{CavalieriMiles}:
\begin{equation}
    \langle C_{\xi_1} \cdots C_{\xi_k} \rangle_{0,n}=n! H_{0,n}^{\bullet} (\xi_i) \, .
    \label{CorrelationFunctions}
\end{equation}
In particular, the two-point function on the sphere reads
\begin{equation}
    \langle C_\phi C_\psi \rangle_{0,n}= [C_e] C_{\phi} C_{\psi}=\lvert [\phi] \rvert \delta_{[\phi],[\psi]} \, .
\label{MetricOnTheSphere}
\end{equation}
As argued previously, the other crucial Feynman diagram in the theory is the one with two incoming states and 
one outgoing state that codes the multiplication in the Frobenius algebra.  Since the states correspond to the conjugacy class sums $C_\mu $, the multiplication equals the multiplication of conjugacy class sums. We have the product
\begin{align}
\mu : \curlyC_n \times \curlyC_n \rightarrow \curlyC_n: (C_{\mu_1},C_{\mu_2}) \mapsto {c_{\mu_1,\mu_2}}^{\mu_3} C_{\mu_3}\, ,
\label{Product}
\end{align}
where we took a basis of conjugacy class sums in the center $\curlyC_n$ of the group 
algebra $\mathbb{C}[S_n]$ and the conjugacy class multiplication has structure constants ${c_{\mu_1,\mu_2}}^{\mu_3}$ also known as the connection coefficients. Determining the connection coefficients explicitly for a large class of partitions is a standard and 
hard problem in combinatorics. There has been a lot of progress in the subject of determining the three-point functions for the Hurwitz theory explicitly, but a full concrete answer is out of reach. 

\subsection{The Partial Permutations and their Extended Algebra}
\label{PartialPermutationsAndExtendedAlgebra}

One of our goals will be to study the exact $n$-dependence of the observables, i.e. of the Hurwitz numbers. A mathematical tool that is ideally suited to that goal is the algebra of orbits of partial permutations \cite{IvanovKerov}. Moreover, at infinite $n$, this algebra will match the algebra of operators of a string theory. In preparation for our discussion of the gauge/string duality in section \ref{H/GW}, we therefore also construct a multiplicative basis for the algebra in which the duality is most naturally expressed.
\subsubsection{The Partial Permutation Algebra}
\label{HurwitzDependenceOnN}
We briefly recall the construction of the algebra of partial partial permutations and its role in the study of the $n$-dependence of the correlators. We define partial permutations $(d,\rho)$ as permutations that pertain to a subset $d$ of $\mathbb{P}_n:=\{1,2,\dots,n\}$. We refer to this subset $d$ as the support of the partial permutation. The permutation $\rho$ acts non-trivially only within this subset but perhaps even so in a smaller subset. To alleviate the notation, we sometimes omit the support and refer to a partial permutation $(d,\rho)$ with the roman letter $r$. Its support is then denoted $d_{r}$ and the associated permutation of $S_n$ is denoted by the corresponding Greek letter, $\rho=\psi(r)$.\footnote{We emphasize that the permutation $\rho$ lives in $S_n$, but acts non-trivially only within the subset $d$.} 

The partial permutations form a semi-group $\mathcal{P}_{n}$. The associated algebra  $\mathcal{B}_n:=\mathbb{C}[\mathcal{P}_{n}]$ is semisimple and equal to a direct sum 
of group algebras summed over subsets of $\mathbb{P}_n$ \cite{IvanovKerov}:
\begin{equation}
	\mathcal{B}_{n}=\bigoplus_{d\subset \mathbb{P}_n} \mathbb{C}[S_{d}] \, .
    \label{ppalgebra}
\end{equation}
There is a surjective morphism $\psi: \mathcal{B}_{n}\to \mathbb{C}[S_{n}]$,
which forgets about the support $d$ and trivially completes a permutation with inactive colours.\footnote{We call the elements of the set $\mathbb{P}_n$ colours and call them inactive if and only if they are mapped to themselves.}
Moreover, there is an action of the symmetric group $S_{n}$ on the algebra of partial permutations $\mathcal{B}_{n}$ defined by:
\begin{equation}
    \sigma \cdot (d,\rho)=(\sigma d, \sigma \rho \sigma^{-1}) \, .
\end{equation}
We denote by $\mathcal{A}_{n}=\mathcal{B}_{n}^{S_{n}}$ the algebra of invariants under this action.
A basis for the invariants $\curlyA_{n}$ is given by the summation over the orbits
of partial permutations
\begin{equation}
    A_{r;n}:=\sum_{(d,\rho)\in [r;n]} (d,\rho)\, ,
\end{equation}
where $r$ is a partial permutation and $[r;n]$ denotes the associated orbit in the algebra $\curlyB_{n}$ under the action of the symmetric group $S_{n}$. Each element in the orbit 
has the same cycle structure as $\rho$ and a support of size $|d_{r}|$, which we refer to as the degree of the orbit $A_{r;n}$. 

\subsubsection{The Extended Algebra}
It is useful to introduce the infinite $n$ limit of the algebra of partial permutations $\curlyA_n$. We denote it $\curlyA_{\infty}$ \cite{IvanovKerov}. We emphasize that the elements of the algebra $\curlyA_{\infty}$ have finite support and its structure constants are $n$-independent. Furthermore, this extended algebra encodes all of the information on the finite algebras $\curlyA_n$. Indeed, one can define a projection morphism $\theta_{n}:\curlyA_{\infty}\to \curlyA_n$ to go from the infinite algebra to the finite one. One then recovers the fact that the structure constants $g_{rs}^{\ \ t}$ of the algebra of invariants $\curlyA_n$ are independent of $n$ if the degrees of the orbits satisfy the constraint $\lvert d_t \rvert \leq n$ \cite{IvanovKerov}. Moreover, the $n$-independence of the structure constants of the algebra $\curlyA_{\infty}$ may be exploited to compute the $n$-dependent structure constants of the physical algebra $\curlyC_n$, as was done in \cite{IvanovKerov,Ashok:2023mow}. 
We recall the following concrete example of a multiplication in $\curlyA_{\infty}$ \cite{Farahat}:
\begin{equation}
    A_{[3]}\cdot A_{[3]}=2 A_{[3^2]}+5 A_{[5]}+8 A_{[2^2]} + 3 A_{[31]}+A_{[3]}+2 A_{[1^3]} \, . \label{ExamplePartialPermutations}
\end{equation}
This equation can be projected to the $\curlyC_n$ algebra using the support-forgetting morphism and terms must then be appropriately gathered. The $n$-dependence of the structure constants of the $\curlyC_n$ algebra arises from the overcompleteness of the basis of partial permutations from the perspective of the conjugacy classes of the group $S_n$, as well as from combinatorial factors. The result is:
\begin{equation}
    C_{[3]}\cdot C_{[3]}=2 C_{[3^2]}+C_{[5]}+8 C_{[2^2]} + (3n-8) C_{[3]}+\frac{n(n-1)(n-2)}{3} C_{e} \, .
    \label{ProductExample}
\end{equation}
The combinatorics captured by the partial permutation algebra is finer than the Hurwitz combinatorics. This is clearly demonstrated in the example calculation of structure constants given in equations (\ref{ExamplePartialPermutations}) and (\ref{ProductExample}) above, or more generally by the fact that the Hurwitz structure constants are linear combinations of the partial permutation algebra structure constants \cite{IvanovKerov}.

Furthermore, we note that the infinite algebra of partial permutations $\curlyA_{\infty}$ may be endowed with the (degenerate) bilinear form $\langle A_{r} \rangle_{0,n} := \langle (\psi_n\circ \theta_n) (A_{r}) \rangle_{0,n}$. The brackets on the right-hand side refer to the correlation functions on $\curlyC_n$. The support-forgetting morphism $\psi_n$ generates a combinatorial factor equal to \cite{IvanovKerov}
\begin{equation}
\psi(A_{r;n})= \binom{m_{1}(\rho)}{m_{1}(r)} C_{\rho;n} \, . \label{CombinatorialFactor}
\end{equation}
We recall that the permutation $\rho$ is the permutation obtained by completing the partial permutation $r$. The bilinear form above can be employed to define generalized Hurwitz numbers
\begin{align}
        H^{\bullet}_{0,n} (r_i) := & \frac{1}{n!}\langle A_{r_1} \cdots A_{r_k} \rangle_{0,n} \label{ppCorrelators} \\
        = & \prod_{i=1}^{k} \binom{m_1(\rho_{i})}{m_1(r_i)} \cdot H_{0,n}^{\bullet}\left(\rho_{1},\dots, \rho_{k}\right)  
        \, .
\end{align}
Note that if the partial permutations $r_i$ are minimal, i.e. their supports $d_{r_i}$ contain no inactive colours, or if they are maximal, i.e. their support is $\mathbb{P}_n$, then the generalized Hurwitz numbers we defined match the standard ones. In general, they differ by a combinatorial factor. 
We also note that the introduction of the algebra of partial permutations $\curlyA_{\infty}$ displaces the $n$-dependence of the correlation functions
from the product to the bilinear form.

Finally, we provide an explicit expression for the Hurwitz numbers in terms of the structure constants of the extended Hurwitz algebra. As shown above, the Hurwitz numbers $H_{0,n}^{\bullet}\left(\rho_{1},\dots, \rho_{k}\right)$ can be identified with their generalized counterparts $H_{0,n}^{\bullet}\left(r_{1},\dots, r_{k}\right)$ for the corresponding minimal partial permutations $r_i$:
\begin{equation}
    \langle A_{r_1}\cdots A_{r_k} \rangle_{0,n}= \langle C_{\rho_1}\cdots C_{\rho_k} \rangle_{0,n}\, .
\end{equation}
The product of orbits of partial permutations on the left-hand-side can then be expanded, revealing the following structure
\begin{equation}
       \langle A_{r_1}\cdots A_{r_k} \rangle_{0,n} = \sum_{s_1,\dots, s_{k-1},l} g_{r_1 r_2}^{\quad s_2} g_{s_2 r_3}^{\quad s_3} \cdots g_{s_{k-2} r_{k-1}}^{\quad \quad \ \ s_{k-1}} g_{s_{k-1} r_k}^{\quad \quad s_k} \langle A_{s_k}\rangle_{0,n} \, .
\end{equation}
The one-point correlator evaluates to $\binom{n}{m_1(s_k)} \delta_{\psi(s_k),e}$
and the resulting equation reads
\begin{equation}
      H_{0,n}^{\bullet}\left(\rho_{1},\dots, \rho_{k}\right) =\frac{1}{n!}\sum_{s_1,\dots, s_{k-1},l} g_{r_1 r_2}^{\quad s_2} g_{s_2 r_3}^{\quad s_3} \cdots g_{s_{k-2} r_{k-1}}^{\quad \quad \ \ s_{k-1}} g_{s_{k-1} r_k}^{\quad \quad 1^l} \binom{n}{l}\, .
      \label{HurwitzEqualsSumOfPartialHurwitz}
\end{equation}
This equation shows explicitly that the algebra of partial permutations $\curlyA_{\infty}$ contains all of the information on Hurwitz theory for 
each finite value of $n$.  
\subsubsection{The Extended Algebra and the Shift Symmetric Functions}
In the previous subsection, we have demonstrated a useful rewriting of both conjugacy class sums and Hurwitz numbers that codes all finite $n$ theories in an $n$-independent extended algebraic structure. We wish to perform the same feat on the side of the character formulas for the correlation functions. 
To that end, we extend Burnside's formula (\ref{PartitionFunctions}) to partial permutations and reshape it into
\begin{equation}
    \langle A_{r_1}\cdots A_{r_k} \rangle_{0,n}=\sum_{\lvert\lambda\rvert=n} \left(\frac{\dim \lambda}{n!}\right)^2 \prod_{i=1}^{k}f_{r_i}(\lambda)\, ,
    \label{Burnside}
\end{equation}
where we defined the generalized character $f_r(\lambda)$:
\begin{equation}
    f_{r}(\lambda):=\binom{n}{\lvert d_r \rvert} \lvert C_{\rho}\rvert \frac{\chi_{\lambda}({\rho})}{\dim \lambda}
\end{equation}
and where $\chi_{\lambda}$ is the character of the representation $\lambda$ of the symmetric group $S_n$. 
A crucial insight hides in the above rewriting: the observation that the asymptotic representation theory of the symmetric group also benefits from studying the characters as functions of the representation instead of the conjugacy class \cite{KerovOlshanski, OkounkovOlshanski}. The function $f_{r}(\lambda)$ is zero when $|d_r|>|\lambda|$ and 
is defined by the inclusion of symmetric groups when $|d_r|<|\lambda|$. As such, the function makes sense for any order of the symmetric group. Viewed as functions of the representations, the characters $f_{r}$ become shift symmetric functions \cite{KerovOlshanski, Meliot}. That is, they are invariant under the shifted action 
of the symmetric group, which permutes the variables  $\lambda_i-i$ related to the lengths $\lambda_i$ of the partition $\lambda$:
\begin{equation}
    \sigma \cdot f(\lambda_1-1,...\lambda_n-n)=f(\lambda_{\sigma(1)}-\sigma(1),...\lambda_{\sigma(n)}-\sigma(n)) \, .
\end{equation}
We consider the algebra of functions $\mathbb{C}^{\mathcal{P}(n)}$ on partitions of $n$. The functions invariant under the shifted action form an algebra, which we denote $\mathbb{C}[\lambda_1,...,\lambda_n]^{*S_n}$. It has a natural filtration by degree. 
The algebra $\Lambda^\ast$ of shift symmetric functions is defined as a projective limit of these finite $n$ algebras. Its elements are sequences of polynomials $f=\{f^{(n)}\in \mathbb{C}[\lambda_1,...,\lambda_n]^{*S_n}\}_{n\in \mathbb{N}_0}$ which are stable under restrictions, meaning $f^{(n+1)}\rvert_{\lambda_{n+1}=0}=f^{(n)}$.

Equation (\ref{Burnside}) suggests a relation between orbits of partial permutations and  shift symmetric functions. This correspondence is provided by the Fourier transform isomorphism on  the algebra $\mathcal{C}_{\infty}$, which can be extended to $\mathcal{A}_{\infty}$ \cite{IvanovKerov}:
\begin{align}
    \phi: \mathcal{A}_{\infty} &\to \mathbb{C}^{\mathcal{P}}:
    A_{r} \mapsto \sum_{\sigma \in [r]} \chi_{\bullet}(\sigma)=f_{r}\, ,
\end{align}
where $\mathcal{P}$ denotes the set of partitions of all integers. 
The image of the extended Fourier morphism is in the algebra of shift symmetric polynomials. 
%
%
%
%
Thus, our observables are shift symmetric functions of the input partitions $\lambda_i$.\footnote{It can be shown that $f_\rho$ also arises as the eigenvalue of an interesting operator $a_\rho$ built out of a sum of permutations of cycle type $\rho$ \cite{KerovOlshanski,Meliot}.}


\subsubsection{A Multiplicative Basis}
In this subsection, we wish to motivate a new choice of basis. If we imagine a two-dimensional topological gravity or string theoretic dual, we can follow the lore that single cycles should correspond to single vertex operators while multi-cycles correspond to a product of vertex operators.\footnote{We warn the reader that we will find corrections to the lore.} However, to implement such a duality on insertions in the path integral, we must have a (shift symmetric) multiplicative basis.

Indeed, suppose we have (gravitational, closed string) vertex operators $\tau_k$ labelled by integers $k \ge 0$ as we are used to in theories of topological gravity (with powers $k$ of the tangent bundle representing descendants). Our observables are their $n$-point correlations functions. On the boundary (i.e. on the gauge theory side), we have Hurwitz numbers labelled by  partitions $\lambda_i$.
Suppose we have a basis $b_\lambda$ of the class functions (or rather, orbits of partial permutations) in which it is true that a class function labelled by a partition equals the product of class functions of the members of the partition:
\begin{equation}
b_{\lambda}=b_{[\lambda_1,\dots,\lambda_q]} =
b_{\lambda_1} \dots b_{\lambda_q} \, .
\end{equation}
Then we could rather label the basis elements by integers and just generate the other basis elements of the vector space through multiplication. We can then map:
\begin{equation}
\tau_k \leftrightarrow b_k \, .
\end{equation}
We already saw that the characters are shift symmetric. We can therefore indeed ask for a multiplicative basis of shift symmetric functions. 

In the ordinary symmetric functions, there are three standard multiplicative bases\cite{Macdonald}: the homogeneous, elementary and power sum polynomials. These can be converted to shift symmetric bases by a change of variable $\lambda_i\to \lambda_i-i$. Of these, only the power sums are naturally related to the monomials labelled by a cyclic permutation. Given the lore that a single cycle maps to a closed string bulk operator, we may suspect that the power sum shift symmetric functions are an appropriate choice of basis. This will turn out to be manifest from the string perspective in section \ref{H/GW}. Therefore, we present this basis in more detail.

\subsubsection{The Completed Cycles}
\label{CompletedCyclesInHurwitz}
A  multiplicative basis of the shift symmetric functions is that spanned by the power sum shift symmetric functions $p_r$. 
There is a linear isomorphism $\phi$ between the central elements of the group algebra and the characters. We can extend it to the partial permutations and shifted symmetric functions.  We can then apply the isomorphism to the power sums to find a multiplicative basis $\overline{A_\mu}$ in the algebra of partial permutations:
\begin{equation}
\overline{A_\mu}= \frac{1}{\prod_i r_i} \phi^{-1}(p_r) \, .
\end{equation}
%
%
We call these combinations $\overline{A_\mu}$ of sums of orbits of partial permutations {\em completed cycles} \cite{OP1}. 

Let us be more explicit. The shifted power sums are defined as \cite{OP1}
\begin{equation}
p_k (\lambda) := \sum_{i=1}^\infty (\lambda_i -i + \frac{1}{2})^k-(-i+\frac{1}{2})^k+ (1-2^{-k}) \zeta(-k) \, .  \label{DefinitionPowerSumPolynomials}
\end{equation}
The second term ensures that the sum is finite (for a finite number of non-zero $\lambda_i$) and stable under the restriction to $\lambda_{n+1}=0$. The third term is a regularization 
of the second term that compensates for the subtraction. Moreover, the $1/2$ shift is arbitrary but it is convenient, as it allows for the realization of an extra $\mathbb{Z}_2$ symmetry \cite{KerovOlshanski}. Finally, we define $p_{r}:=\prod_{i}p_{r_i}$.

The highest order term in the completed orbit $\bar{A}_{\mu}$ equals \cite{KerovOlshanski}:
\begin{equation}
    \bar{A}_{r}=A_{r}+\ \text{terms of lower partition degree.}
\end{equation}
The completed cycles $\bar{A_{r}}$ can be thought of as a correction to the standard lore we mentioned above and according to which vertex operators are dual to single cycles. 
Below, we have indicated the coefficients of the orbits of partial permutations in the completed cycles, up to order six. To compute these coefficients, we calculated tables of symmetric shifted power sum polynomials, tables of extended and normalized characters and solved for the change-of-basis matrix. Indeed, by virtue of the map between orbits and extended characters, the elements of this matrix are the coefficients of orbits in completed cycles.
We find:
\begin{align}
    (\bar{1}) &=(1)-\frac{1}{24}()\\
    (\bar{2}) &= (2) \nonumber \\
    (\bar{3}) &= (3)+(1,1)+\frac{1}{12}\cdot (1)+\frac{7}{2880}\cdot () \nonumber\\
    (\bar{4}) &= (4)+2\cdot (2,1)+\frac{5}{4}\cdot (2) \nonumber\\
    (\bar{5}) &= (5)+3\cdot (3,1)+4\cdot (2,2)+\frac{11}{2}\cdot (3)+4\cdot (1^3) +\frac{3}{2}\cdot (1,1)+\frac{1}{80}\cdot (1)-\frac{31}{40320}\cdot () \nonumber\\
    (\bar{6}) &= (6) + 4\cdot (4,1) + 6 \cdot (3,2) +\frac{95}{6} (4) + 10\cdot (2,1,1) +\frac{35}{3}\cdot (2,1)+ \frac{91}{48}\cdot (2) \, . \nonumber
\end{align}
In this formula, we use the notation $r \equiv A_r$ for conjugacy orbits of partial permutations.\footnote{ 
In  Appendix \ref{AlternativeCompletedCycles}, we provide alternative completed cycles corresponding to the power sum shift symmetric polynomials without the $\zeta$-function term in equation (\ref{DefinitionPowerSumPolynomials}).}
Finally, let us point out that these completed cycles have geometric interpretations as Hurwitz numbers for degenerate coverings, see \cite{SSZ} and \cite{EskinOkounkov}.\footnote{The completed cycles are associated to a compactification of the moduli space of Riemann surfaces in section \ref{GaugeStringDuality}.}
\section{Topological Symmetric Orbifold Theories}
\label{SecondQuantizedtopological quantum field theory}
In this section, we show that the Hurwitz theory is the simplest special case of a symmetric orbifold topological field theory. We define an analog of the extended Hurwitz theory in the general case and exploit the combinatorics it defines to study the $n$-dependence of the correlation functions of topological symmetric orbifold theories. 
Subsection \ref{SymmetricOrbifoldTheory} is a review of mathematics literature. Starting from subsection \ref{ExtendedSymmetricOrbifoldAlgebra} we are on new terrain created by importing the extended algebra  ideas related to partial permutations into the realm of symmetric orbifold algebras. Subsection \ref{HierarchyOfCombinatorics} is original as well and subsection \ref{SpecialCases} illustrates how the literature has only discussed very special cases of our construction in those subsections. 
\subsection{The Symmetric Orbifold Theory}
\label{SymmetricOrbifoldTheory}
Consider a two-dimensional conformal field theory $X$. Through the state-operator correspondence, its state space defines an operator algebra $A$. We denote its elements by lowercase letters $a_i$. The symmetric orbifold theory $\text{Sym}_{n}(X):=X^n/S_n$ is defined as follows. First, one allows for twisted boundary conditions in the  theory $X^{\otimes n}$ consisting of $n$ copies of the theory $X$
\begin{equation}
	a_{I}:=\bigotimes_{i=1}^{n}\, a_i , \, a_{I}(\theta+2\pi)=a_{\sigma(I)} \, .
 \label{TwistedSector}
\end{equation}
Each choice of non-trivial permutation $\sigma$ gives rise to a twisted sector and the preliminary space of operators is 
\begin{equation}
	A\{S_n\}=\bigoplus_{\pi\in S_n} A^{\otimes \langle \pi \rangle \backslash [n]}\pi  \, ,
\end{equation} 
where $[n]$ is a more compact notation for $\mathbb{P}_n$ and $\langle \pi \rangle \backslash [n]$ is the space of orbits of $\mathbb{P}_n$ under the action of the permutation $\pi$. Each orbit corresponds to a cycle of $\pi$ and one should think of elements in the unprojected space $A\{S_n\}$ as grouping together the element of the algebra $A$ and its associated cycle. If we denote the decomposition of 
the permutation $\pi$ into cycles as $\pi=\pi_1\cdots \pi_k$, then we have:
\begin{equation}
    a\pi \equiv (a_1 \pi_1) \dots (a_k \pi_k) \, .
\end{equation}
The second step in the definition of the symmetric orbifold theory consists in projecting on gauge invariant states
\begin{equation}
    A^{[n]}:=A\{S_n\}^{S_n} \, .
\end{equation}
The  sectors are in one-to-one correspondence with the conjugacy classes of the symmetric group. A basis for elements of the symmetric orbifold algebra $A^{[n]}$ is given by the operators
\begin{equation}
    O_{[\pi];n}(a):=\frac{\lvert [\pi] \rvert}{n!} \sum_{\sigma\in S_n} \sigma^*(a)\, \sigma \pi \sigma^{-1}\, ,
    \label{ExtendedOperatorBasisDuplicate}
\end{equation}
where the symmetric group acts on the element $a$ by dragging its tensor factors associated to given orbits of $\pi$ to the corresponding orbits of $\sigma \pi \sigma^{-1}$. 
For example, for $n=3$, if $a\pi=a_1(12)\, a_2(3)$, then $O_{[\pi];3}(a)=a_1(12)\, a_2(3)+a_1(13)\, a_2(2)+a_1(23)\, a_2(1)$. The normalization factor in front of the sum is chosen such that for a trivial seed theory, the operators $O_{[\pi];n}(a)$ reduce to the conjugacy class sums $C_{[\pi]}$ of the symmetric group $S_n$. In fact, as we show in section (\ref{SpecialCases}), the symmetric orbifold of a trivial seed theory reduces to Hurwitz theory.

\begin{figure}[ht]
	\centering
	\includegraphics[height=8cm]{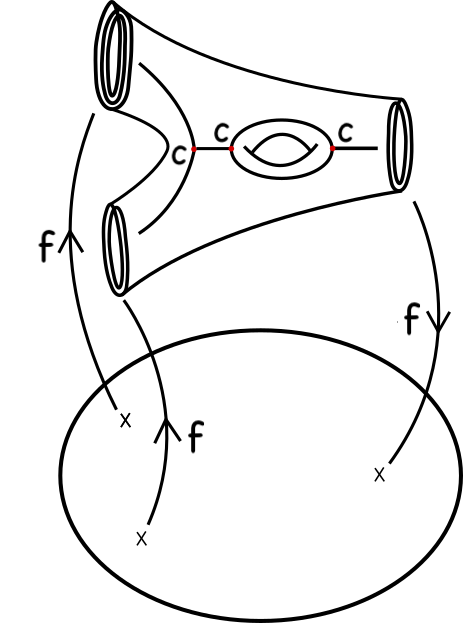}
	\caption{The worldsheet (at the bottom) has twist operator insertions that become ordinary operators on the covering surface (at the top). The latter multiply ordinarily.}
 \label{MultiplicationAndCovering}
\end{figure}

Correlation functions in the symmetric orbifold theory are notoriously hard to compute. One strategy consists in performing the path integral explicitly in the covering space of the manifold with the twist insertions \cite{Lunin:2000yv,Lunin:2001pw}. The situation simplifies drastically in the topological setting. The space of operators $A$ is then endowed with the structure of a Frobenius algebra. Let $T$ be the Frobenius one-form on $A$ and $\eta^A$ the non-degenerate bilinear form it induces. The correlation functions are encapsulated in the product on the derived algebra $A^{[n]}$. 
Furthermore, it was shown in \cite{Kaufmann} that a unique product can be defined on this vector space $A^{[n]}$, up to discrete torsion, given the seed product. 
This product must therefore coincide with the more explicit description of \cite{LS2} reviewed in \cite{Li:2020zwo,Ashok:2023kkd} in detail.
We very briefly recall the formula for the product $m$ of two operators $a \pi$ and $b \rho$:
\begin{equation}
\label{DefOfMultiplication}
m_{\pi,\rho}(a \otimes b) = f_{\langle \pi,\rho \rangle,\langle \pi \rho \rangle}( f^{\pi,\langle \pi,\rho \rangle}(a) f^{\rho,\langle \pi,\rho \rangle}(b) e^{g(\pi,\rho)})
\end{equation}
where the maps $f^{\pi,\langle \pi,\rho\rangle}$ multiply together the operators that sit in the same orbit of the second group in the upper index, $e$ is the Euler class of the Frobenius algebra and  $g(\pi,\rho)$ is the genus of the covering surface determined by the permutations $\pi$ and $\rho$.\footnote{That the graph defect $g(\pi,\rho)$ agrees with the genus $g$ of the covering surface is proven in \cite{Barmeier:2010uze} Lemmas 11 and 12.}
The final application $f_{\langle \pi,\rho\rangle,\langle \pi \rho \rangle}$ comultiplies the resulting operator into the orbits of the outgoing permutation $\pi \rho$. 
Instead of reviewing the product \cite{LS2} once more \cite{Li:2020zwo,Ashok:2023kkd}, we propose a picture for the interpretation of the various ingredients. The product corresponds to the pants diagram in Figure \ref{MultiplicationAndCovering}. 
Two operators multiply into one. By holonomy considerations, the product $\pi \rho$ of the permutations of the ingoing operators equals the permutation associated to the outgoing operator. On the worldsheet, there are twisted operator insertions that are multi-valued. For a twisted sector, the operator $a_I$ is not mapped to itself after we change its argument by $2 \pi$ -- see equation (\ref{TwistedSector}). The operators become single valued on the cover. This fact is incorporated in the definition of the space $A \{ S_n \}$. Thus, to augment intuition on the product of operators, it is useful to first lift to the cover where the operators are local. 
Moving the operators to the cover is implemented by the maps $f$ inside the parentheses. On the cover, operators multiply as local operators. This introduces cubic vertices corresponding to the structure constants of the seed Frobenius algebra. 
The covering map may have loops counted by the genus or graph defect $g(\pi,\rho)$. In that case, we must work with the dual cubic vertices and multiply in handle operators $e$ of the seed Frobenius algebra. This again is part of the definition of the multiplication.
The final map $f$ with lower indices is used to redescend to the original worldsheet. This involves the co-multiplication in the seed Frobenius algebra. 
The comultiplication map $\Delta_*^{(m)}$  of order $m$ sends a basis vector $e_l$ to
\begin{equation}
    \Delta_*^{(m)}(e_l)=c^{k_1 k_2 j_3} c_{j_3}^{\ k_3 j_4} \dots c_{j_m}^{\ k_m j_{m+1}} \eta_{l,j_m+1} e_{k_1}\otimes \dots \otimes e_{k_m} \, .
\end{equation}
See \cite{LS2,Kaufmann,Li:2020zwo,Ashok:2023kkd} for more details.

Finally, we recall that the Frobenius one-form $T$ on the seed algebra induces a Frobenius structure on the symmetric orbifold algebra $A^{[n]}$ \cite{LS2}
\begin{equation}
\label{OneFormOnSymOrbifold}
    T[O_{[\pi]}(a)]=
    \begin{cases}
        T(a):=T(a_1)\cdot \dots \cdot T(a_n)\  \text{if}\ \pi=e\, .\\
        \enskip 0\ \text{otherwise}\, .
    \end{cases}
\end{equation}

\subsection{The Extended Symmetric Orbifold Algebra}
\label{ExtendedSymmetricOrbifoldAlgebra}
Our analysis of the Hurwitz topological quantum field theory and its string dual rests on an extension of the finite $n$ operator algebra. In this subsection, we define an extended operator algebra for the symmetric orbifold theory. We first define the grand canonical space
\begin{equation}
    \mathcal{B}_{\infty}(A)=\bigoplus_{d\subset \mathbb{P}_\infty} A\{S_d\}\, .
    \label{UnprojectedSpace}
\end{equation}
Elements of the algebra $\mathcal{B}_{\infty}(A)$ are of the form $a p$ where $p=(d_p,\pi)$ is a partial permutation and $a\in A^{\otimes \langle \pi \rangle \backslash [d_p]} $, with $\langle \pi \rangle \backslash [d_p]$ denoting the space of orbits of the set $d_p$ under the action of the permutation $\pi$. The product on $\mathcal{B}_{\infty}(A)$ is defined as follows:
\begin{equation}
    a p\cdot b r=m_{p,r}(\psi_{d_{p},d_{p}\cup d_{r}}(a)\otimes  \psi_{d_{r},d_{p}\cup d_{r}}(b))\, ,
    \label{MultiplicationOnAlgebra}
\end{equation}
where $\psi_{d_1,d_2}$ augments the support of a partial permutation to $d_1\cup d_2$ and inserts an identity operator for every unmarked colour: 
\begin{align}
     \psi_{d_1,d_2}: \, & A\{S_{d_1}\} \to A\{S_{d_1\cup d_2}\} \\
     & a(d_1,\pi) \mapsto a\cup 1^{d_2\backslash d_1} (d_1 \cup d_2, \pi)\, .
\end{align}
The element $a\cup 1^{d_2\backslash d_1}$ denotes the element of the algebra $A^{\langle \pi \rangle \backslash [d_1\cup d_2]}$ obtained by inserting ones in the free slots of the new support $d_1\cup d_2$. The multiplication $m$ in equation (\ref{MultiplicationOnAlgebra}) proceeds as in equation (\ref{DefOfMultiplication}), replacing permutations with partial permutations in all formulas. We note that we sometimes abuse the above notation and denote by $\psi_{d_1,d_2}$ the part of the map that acts only on $A^{\langle \pi \rangle \backslash [d_p]}$, as in equation (\ref{MultiplicationOnAlgebra}).

We now project on the gauge-invariant states and define the algebra  $\curlyA_{\infty}(A):=\mathcal{B}_{\infty}(A)^{S_{\infty}}$.\footnote{We define the $n\to \infty$ limit of the symmetric group $S_n$ as the group of permutations of the set $\mathbb{N}_0$ of strictly positive natural numbers that only shuffle a finite number of elements.}
We refer to this algebra as the extended operator algebra of the theory $A$. The state space associated to the theory is a Fock space labelled by the elements of the seed theory \cite{LS2,Dijkgraaf:1996xw}. We may therefore think of this extended algebra as a grand canonical approach to the symmetric orbifold theory.
Furthermore, the sum (\ref{UnprojectedSpace}) can be truncated to define an $n$-extended operator algebra $\curlyA_n(A)$ at finite $n$.\footnote{Strictly speaking, one first defines the algebra $\curlyA_n(A)$ and then the algebra $\curlyA_{\infty}(A)$ as its projective limit with respect to a projection morphism.}
This is the appropriate abstract setting for the computations performed in \cite{Ashok:2023kkd}, as we shall prove below. The basis operators of this $n$-extended algebra are denoted $O_{[p];n}(a)$ and defined as
\begin{equation}
    O_{[p];n}(a):=\frac{\lvert [p;n] \rvert}{n!} \sum_{\sigma\in S_n} \sigma^*(a)\, \sigma p \sigma^{-1}\, ,
    \label{ExtendedOperatorBasis}
\end{equation}
where the symmetric group acts on partial permutations by dragging the tensor 
factors of $a$ that are associated to a given orbit of $p$ to the corresponding 
orbit of $\sigma p \sigma^{-1}$. The $n$-dependence of the normalization factor $\curlyN_p=\lvert [p;n] \rvert/n!$ cancels out with that of the $n$-dependent number of identical partial permutations that arise in the sum, such that the large $n$ limit of the operators is well-defined. Moreover, the normalization factor is chosen such that our operators reduce to the orbits of partial permutations $A_{p;n}$ when the seed theory $A$ is trivial. In fact, for a trivial seed algebra, the extended symmetric orbifold algebra $\curlyA_{\infty}(A)$ reduces to the extended Hurwitz algebra $\curlyA_{\infty}$.

\subsection{A Hierarchy of Combinatorics}
\label{HierarchyOfCombinatorics}
The approach we followed to study the $n$-dependence of observables in the Hurwitz theory can be generalized to the symmetric orbifold case. It is summarized in the diagram below.
\begin{figure}[H]
    \centering
    \includegraphics[width=4.5cm]{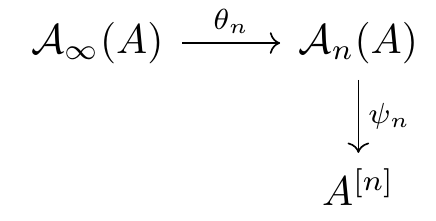}
    \vspace{-1.5mm}
\end{figure}
\noindent The first step is to demonstrate the $n$-independence of the structure constants of $\curlyA_{n}(A)$. The second is to compute the structure constants in the extended algebra $\curlyA_{\infty}(A)$. The third is to push the structure constants of the algebra $\curlyA_{n}(A)$ to the algebra of the symmetric orbifold theory $A^{[n]}$.
\subsubsection{Finite and Infinite Structure Constants}
We define the projection morphism
\begin{align}
    \hat \theta_{n}: \curlyB_{\infty}(A) &\to \curlyB_n(A)
                     \nonumber \\
                  a p &\mapsto 
                  \begin{cases}
                    a p\ \text{if}\ d_{p} \subset \mathcal{P}_n\, .\\
                     \hspace{0.7mm} 0\ \text{otherwise}.
                  \end{cases}
\end{align}
The morphism $\hat \theta_{n}$ commutes with the action of the symmetric group $S_n$. It therefore induces a morphism $\theta_{n}$ by restriction to the gauge-invariant subspace: 
\begin{align}
    \theta_{n}:\, &\curlyA_{\infty}(A) \to \curlyA_n(A) \nonumber \\
    & O_{[r]}(a) \mapsto
        \begin{cases}
            O_{[r;n]}(a)\ \text{if}\ \lvert d_{r} \rvert \leq n\, .\\
            \hspace{1.3mm} 0\ \text{otherwise}.  
        \end{cases}
\end{align}
Acting with the projection morphism $\theta_n$ on a product of operators in the infinite extended algebra $\curlyA_{\infty}(A)$, we find that its $n$-independent structure constants $g_{rs}^{\ \ t}(i,j;k)$ match the structure constants $(g_n)_{rs}^{\ \ t}(i,j;k)$ of the $n$-extended algebra $\curlyA_n(A)$. The latter must therefore be $n$-independent. As in the Hurwitz case, this $n$-independence comes with the caveat that if $ \lvert d_{t} \rvert > n$,  the finite $n$ structure constants $(g_n)_{rs}^{\ \ t}(i,j;k)$ vanish regardless of whether the constants $g_{rs}^{\ \ t}(i,j;k)$ do.
\subsubsection{Combinatorics of the Extended Symmetric Orbifold}
We now compute the structure constants in the extended algebra $\mathcal{A}_{\infty}(A)$. To alleviate the expressions, we denote the action of conjugation by the group element $g$ by $\sigma_g$. We have:
\begin{align}
\frac{O_{[p]}(a)}{\curlyN_{[p]}}\cdot \frac{O_{[r]}(b)}{\curlyN_{[r]}} &= 
\sum_{\alpha,\beta \in S_n}  m_{\sigma_{\alpha} p,\sigma_{\beta} r}\Big(\psi_{d_{\sigma_{\alpha} p}\cup d_{\sigma_{\beta} r}}(\alpha^{\ast}(a))\otimes \psi_{d_{\sigma_{\alpha} p}\cup d_{\sigma_{\beta} r}}(\beta^{\ast}(b))\Big)\ \alpha p \alpha^{-1}  \beta r \beta^{-1}
\nonumber \\
&= \sum_{\beta,\gamma \in S_n}  m_{\sigma_{\beta} \sigma_{\gamma} p, \sigma_{\beta} r} \Big(\psi_{d_{\sigma_{\beta} \sigma_{\gamma} p}\cup d_{\sigma_{\beta} r}}(\beta^{\ast}(\gamma^{\ast}(a))\otimes \psi_{d_{\sigma_{\beta} \sigma_{\gamma} p}\cup d_{\sigma_{\beta} r}}(\beta^{\ast}(b))\Big) \nonumber \\
& \qquad \qquad \qquad \qquad \qquad \qquad \qquad \qquad \qquad \qquad \qquad \qquad  \beta (\gamma p \gamma^{-1} r) \beta^{-1} \nonumber \\
&=\sum_{\gamma \in S_n} \sum_{\beta \in S_n} \beta^* \left(m_{\sigma_{\gamma}p,r} (\psi_{d_{\gamma p {\gamma}^{-1}}\cup d_{r}}(\gamma^{\ast}(a)) \otimes \psi_{d_{\gamma p \gamma^{-1}}\cup d_{r}}(b))\right) \beta (\gamma p \gamma^{-1} r) \beta^{-1} \, .
\nonumber
\end{align}
We conclude that
\begin{equation}
    O_{[p]}(a) O_{[r]}(b) =\sum_{\gamma\in S_n}\frac{\curlyN_{[p]} \curlyN_{[r]}}{\curlyN_{[\gamma p \gamma^{-1} r]}}\cdot O_{[\gamma p \gamma^{-1} r]}\Big(m_{\gamma p \gamma^{-1},r} (\psi_{d_{\sigma_{\gamma}p}\cup d_{r}}(\gamma^{\ast} (a)) \otimes \psi_{d_{\sigma_{\gamma}p}\cup d_{r}}(b))\Big)
    \label{ProductOfBasisOperators} \, .
\end{equation}
In going from the first to the second line, we have performed the change of variable $\gamma=\beta^{-1}\alpha$. In going from the second to the third line, we have used that the product in $\curlyB_{\infty}(A)$ is $S_n$-equivariant, which follows from the $S_n$-equivariance of the product in $A\{S_n\}$:\footnote{Proposition 2.13  in \cite{LS2}.}
\begin{equation}
    \sigma^*(a p\cdot b r)= \sigma^*(a p) \cdot \sigma^*(b r)\, .
\end{equation}
In principle, one can regroup terms for which the operator on the right-hand side of equation (\ref{ProductOfBasisOperators}) are identical regardless of the choice of seed algebra. 
The resulting equation can schematically be written  as
\begin{equation}
    O_{[p]}(a) O_{[r]}(b)=\sum_{s,c} k_{pr}^{\ \ s}(c)\, O_{[s]}(c)\, ,
    \label{SchematicProductInExtendedAlgebra}
\end{equation}
where the symbol $c$ denotes the various arguments of $O_{\gamma p \gamma^{-1} r}$ in equation (\ref{ProductOfBasisOperators}).\footnote{We treat $c$ as a formal product and do not evaluate it in a specific seed algebra.} The coefficients $k_{pr}^{\ \ s}(c)$ are equal to sums of the ratios of normalization factors. They are new combinatorial numbers independent of the seed algebra $A$ and we refer to them as reduced structure constants. They are not to be confused with the structure constants $g_{pr}^{\ \ s}(a,b;c)$ of the algebra $\curlyA_{\infty}(A)$ which depend on the seed algebra through the product of operators hidden in the argument $c$.\footnote{The structure constants $g_{pr}^{\ \ s}(a,b;c)$ may be written as sums of structure constants of the seed algebra weighted by the reduced structure constants.}
Furthermore, for a trivial seed theory, all operators on the right-hand side of equation (\ref{ProductOfBasisOperators}) are identical and we recover the structure constants of the algebra of orbits of partial permutations
 \begin{equation}
     g_{pr}^{\ \ s}=\sum_{c} k_{pr}^{\ \ s}(c)\, .
     \label{SymmetricOrbifoldConstantsArePartitions}
 \end{equation}

Let us analyze these new combinatorial numbers. Firstly, we will show that the combinatorics of single cycles reduces to that of the partial permutation algebra $\curlyA_{\infty}$. Then, we will explain how the extended symmetric orbifold algebra $\curlyA_{\infty}(A)$ provides a refinement of the partial permutation combinatorics.

Denote by $s_k=(\mathbb{P}_k,\alpha_k)$ the partial permutation corresponding to the single cycle $\alpha_k=(1\dots k)$. We compute the terms
\begin{equation}
    O_{[\sigma_{\gamma}(s_i) s_{j}]}\Big(m_{\sigma_{\gamma}(s_i),s_{j}}\left(\psi_{\sigma_{\gamma}(s_i)\cup s_{j}}(\gamma^*(a))\otimes \psi_{\sigma_{\gamma}(s_i)\cup s_{j}}(b)\right)\Big) \, .
\end{equation}
The argument of the multiplication $m_{\sigma_{\gamma}(s_i),s_{j}}$ is given by $(a\otimes 1^{d_{s_j}\backslash d_{\sigma_{\gamma}(s_i)}})\otimes (b\otimes 1^{ d_{\sigma_{\gamma}(s_i)}\backslash d_{s_j}})$ with ones inserted in the slots introduced by the morphism $\psi$. If the two cycles $\sigma_{\gamma}s_i$ and $s_j$ have no overlap, then acting with the functions $f^{\sigma_{\gamma}s_i,\langle \sigma_{\gamma}s_i, s_{j} \rangle}$ and $f^{s_j,\langle \sigma_{\gamma}s_i, s_{j} \rangle}$ respectively and  taking the product yields $a\otimes b$. If they do have an overlap, then we find $abe^{g}$, where the genus $g$ is determined by the overlap. 
Finally, one has to act with the function $f_{\langle \sigma_{\gamma}s_i, s_{j} \rangle,\sigma_{\gamma}(s_i) s_{j}}$. If the cycles have no overlap, this function acts trivially. If they do have an overlap, then we end up with the comultiplication $\Delta_{*}^{(m)}(abe^{g})$, where $m$ is the number of disconnected cycles in $\sigma_{\gamma}(s_i) s_{j}$. 

Therefore, in the case of a product of single cycles, the terms on the right-hand side of equation (\ref{ProductOfBasisOperators}) that have a given cycle structure are marked with the same operator. Gathering them, we find that the reduced structure constants of single cycles $k_{s_i s_j}^{\quad t}(c)$ match the structure constants $g_{s_i s_j}^{\quad t}$ of the algebra of partial permutations $\curlyA_{\infty}$. We write:
\begin{equation}
    O_{[s_{i}]}(a) O_{[s_{i}]}(b)= g_{s_{i}s_{j}}^{\quad t} O_{[t]}(c)\, ,
    \label{ProductOfSingleCycles}
\end{equation}
where the arguments $c$ are elements of the algebras $A^{\tau \backslash [d_t]}$. The single cycle operators $O_{[s_i]}(a)$ thus constitute a special subclass of the extended algebra $\curlyA_{\infty}(A)$.
We note that there is no contradiction between the statement that single cycles in the algebra $\curlyA_{\infty}(A)$ have the same combinatorics as the algebra $\curlyA_{\infty}$ and the fact that the combinatorics of the former is more refined. Indeed, although single cycles do generate the whole algebra, our formula (\ref{ProductOfSingleCycles}) does not provide sufficient information to compute the product of a single cycle operator with a multi-cycle operator. As we illustrate below, it is from the multi-cycle operators with cycles of identical length that the refinement of the combinatorics arises.

In general, the operators $O_{[t]}(c)$ on the right-hand side of equation (\ref{ProductOfBasisOperators}) do not coincide for the different values of $\gamma$ 
that satisfy the condition $\gamma p \gamma^{-1} r \in [t]$. This is due to the possible presence of cycles of the same length in the operators we multiply. Indeed, such cycles can be exchanged under conjugation and lead to different products of operators $c$ in the argument of $O_{[s]}$. Consider the following example: 
\begin{align}
    \label{NonTrivialRefinementExample}
    O_{2^2}(a,b) \cdot O_2(c)=\, &3O_{2^3}(a,b,c)+\frac{3}{2}\Big(O_{3,2}(ac,b)+O_{3,2}(bc,a)\Big)\\
    & +2O_4(abc)+\frac{1}{2}\Big(O_{2,1^2}(a,\Delta_{*}^{(2)}(bc))+O_{2,1^2}(b,\Delta_{*}^{(2)}(ac))\Big)\, .
    \nonumber
\end{align}
%
%
We see explicitly that the symmetrization gives rise to different operator arguments corresponding to the choices of elements that get multiplied. This implies that the 
extended symmetric orbifold algebra $\curlyA_{\infty}(A)$ defines a more refined 
combinatorics than the algebra of partial permutations $\curlyA_{\infty}$. In general, 
the product of operators $O_{[p]}(a)$ gives rise to multiple operators associated with 
the same partial permutation orbit and the reduced structure constants $k_{pr}^{\ \ s}(c)$ 
are fractions of the structure constants of the algebra $\curlyA_{\infty}$ (\ref{SymmetricOrbifoldConstantsArePartitions}). Furthermore, symmetry considerations suggest that all structure constants $k_{pr}^{\ \ s}(c)$ associated with the same partial permutations $p, r$ and $s$ are equal. In other words, the reduced structure constants are obtained by dividing the structure constants $g_{pr}^{\ \ s}$ by a symmetrization factor defined as the number of different ways to multiply the operators $a$ and $b$ that is consistent with the structure of the orbits of partial permutation. Equation (\ref{NonTrivialRefinementExample}) provides an explicit example. 

Importantly, we therefore have a {\em hierarchy of combinatorics}. The Hurwitz combinatorics is refined by the partial permutation algebra \cite{IvanovKerov} as discussed in subsection \ref{HurwitzDependenceOnN}. In turn, the partial permutation algebra is refined by the combinatorics of the extended symmetric orbifold algebra as demonstrated above.\footnote{One can refer to the latter combinatorics as a marked partial permutation combinatorics as it originates from marking cycles (including of length one) of partial permutations with operators. The combinatorics depends on the dimension of the seed algebra $A$ and may simplify under the assumption of an infinite dimension for the latter.}

\subsubsection{Combinatorics of the Symmetric Orbifold}
We now introduce a support-forgetting homomorphism. This allows us to push the structure constants of the algebra $\curlyA_{n}(A)$ to the algebra of the theory $A^{[n]}$. The support-forgetting morphism is induced from the map $\psi_{d_1,d_2}$ as follows. Take $d_2=\mathbb{P}_{n}$ as target and sum over the $d_1=\mathbb{P}_{k}$ for $k\leq n$. Then, project on the $S_n$-invariant states. The resulting map is a morphism and reads 
\begin{align}
     \psi_n: \, & \curlyA_{n}(A) \to A^{[n]} \nonumber \\
     & O_{[r];n}(a)\mapsto \binom{m_1(\rho)}{m_1(r)} O_{[\rho]}(a\cup 1^{n-\lvert d_{r}\rvert })\, .
\end{align}
To illustrate the method, we apply it to a concrete example. The application of the support-forgetting homomorphism to equation (\ref{NonTrivialRefinementExample}) yields
\begin{align}
    O_{2^2}(a,b)\cdot O_2(c)=\, &3O_{2^3}(a,b,c)+\frac{3}{2}\Big(O_{3,2}(ac,b)+O_{3,2}(bc,a)\Big)\\
    & +2O_4(abc)+\frac{(n-2)(n-3)}{4}\Big(O_{2,1^2}(a,\Delta_{*}^{(2)}(bc))+O_{2,1^2}(b,\Delta_{*}^{(2)}(ac))\Big)\, .
    \nonumber
\end{align}
To derive the structure constants of $A^{[n]}$, one then evaluates the products in the seed algebra $A$ and gathers terms, as for the extended algebra $\curlyA_{n}(A)$. This procedure has to be carried out on a case-by-case basis. Note moreover that the reduced structure constants of $A^{[n]}$ are fractions of the structure constants of the Hurwitz algebra $\curlyC_n$. The algebra of conjugacy class sums is refined by the symmetric orbifold combinatorics and the new combinatorial numbers can be thought of as a generalization of Hurwitz numbers.

To conclude, we stress again that the extended algebra of invariants $\curlyA_{\infty}(A)$ encodes all of the combinatorial information on the topological symmetric orbifold theory at finite $n$ for all values of $n$. Indeed, the formula (\ref{HurwitzEqualsSumOfPartialHurwitz}) can be generalized to express the correlation functions (\ref{OneFormOnSymOrbifold}) at finite $n$ in terms of the structure constants of the algebra of invariants $\curlyA_{\infty}(A)$ and the one-point function of the seed theory 
\begin{align}
      \langle O_{\rho_1}(a_1) \cdots  O_{\rho_k}(a_k) \rangle_n &= \left( \prod_{i=1}^k \frac{1}{\binom{m_1(\rho_i)}{m_1(r_i)}}\right) \sum_{\substack{s_1,\dots, s_{k-1}, l \\ b_2,\dots, b_k}} g_{r_1 r_2}^{\quad s_2}(a_1,a_2;b_2)  \cdots \, g_{s_{k-1} r_{k}}^{\quad \quad \ 1^l}(b_{n-1},a_k; b_k) \nonumber \\
      & \hspace{20em} \cdot  \binom{n}{l}\, T(b_k) \, ,
\end{align}
where the $r_i$'s are the minimal partial permutations that give rise to the permutations $\rho_i$.
\subsection{A Few Special Cases}
\label{SpecialCases}
In this subsection we consider a number of special cases to reconnect to previous literature and to provide the reader with simple examples of the generic construction. Firstly, for the trivial seed Frobenius algebra $A=\mathbb{C}.1$, we have the metric $\eta_{11}=1$ and the Euler class $e=1$. 
This special case equals the Hurwitz theory.  

A second special case is where we have a graded Frobenius algebra $A$ with grades from $-d$ to $d$ and $d$ non-zero and where the Euler class $e$ equals zero, $e=0$.\footnote{An example is the cohomology ring on $T^4$ equipped with integration. For details on the graded Frobenius algebra and its relation to cohomology see e.g. \cite{LS2}. In physics, the graded Frobenius algebra corresponds to the Ramond-Ramond ground state sector of a compact ${\cal N}=2$ superconformal field theory in two dimensions graded by R-charge.} 
Then, the graded algebra $\text{grad} (\mathbb{C}[S_n])$ is a quotient of the graded algebra $A^{[n]}$ by the ideal of operators of non-zero charge (on top of the charge of the ground state in that sector).\footnote{Here, the degree of a permutation in $\text{grad} (\mathbb{C}[S_n])$ equals the lengths of cycles minus one, summed over all cycles, while in the algebra $A^{[n]}$ the degree is the sum of the degrees of the operators. How these are related is discussed in \cite{Li:2020zwo}.} This effectively eliminates all components of operators $a \in A$ not along the identity.\footnote{See Theorem 1.1 in \cite{LQW3} for a concrete example.} It gives rise to the degree-preserving product of permutations or pure joining without the possibility of having loops in the covering surface \cite{Li:2020zwo,Ashok:2023mow}. This is a consequence of the Riemann-Hurwitz theorem and the conservation of degree.
The structure constants of this graded theory are $n$-independent
\cite{Farahat}.\footnote{There are only tree diagrams in this theory and in that sense it is similar to an (interacting) planar Yang-Mills theory. However, note that here there we took no limit in $n$ and that the interaction strength is given by $g_s=1/\sqrt{n}$ rather than the 't Hooft coupling.} This implies the $n$-independence of the cup product structure constants of Hilbert schemes of smooth projective surfaces \cite{LQW2}.
%

Thirdly, there is a description of a particular algebra, namely the cohomology of the symmetric product of ALE spaces, which fits into this framework \cite{Wang} \footnote{Since the target space is non-compact, it provides an algebra of operators rather than a full-fledged topological quantum field theory.} as do other cohomology theories of Hilbert schemes of surfaces. For further interesting examples related to symmetric orbifolds of minimal models via supersymmetric Landau-Ginzburg models, see  \cite{Kaufmann}.

%
%

\section{The Topological Gauge/String Duality}
\label{H/GW}
\label{GaugeStringDuality}
In this section, we briefly review how the Gromov-Witten invariants for the manifold $\mathbb{P}^1$ were calculated and the equality of these invariants with Hurwitz numbers for completed cycles \cite{OP1}. It provides a very precise gauge/string duality, rather unique in its kind.\footnote{See \cite{Lerche:2023wkj} for an exploration in the context of the Hilbert scheme of points on the K3 manifold.}  We draw lessons from this guiding example and frame its proper interpretation using the extended algebra previously introduced.
In the physics literature, this model appeared in the study of two-dimensional gravity in its topological guise \cite{Witten:1989ig} coupled to topological matter \cite{Witten:1989ig,Dijkgraaf:1990nc}. Strong constraints on the amplitudes of primaries and descendants were determined \cite{Witten:1989ig,Dijkgraaf:1990nc,Eguchi:1997jd}. The systematic calculation of all amplitudes was pushed considerably further in \cite{OP1}. We provide intuition for the calculation in our brief review below. The integrable hierarchy of the amplitudes was further clarified in \cite{Okounkov:2003rf}. 

Subsections \ref{GWOnTheSphere} and \ref{OnePointAndCompletedCycles} are only original as a guide to the mathematical literature for physicists. In subsection \ref{GSCorrespondence} we take an explicit step beyond our main reference \cite{OP1} in that we stress that the sum over world sheet instantons is necessary to equip the infinite algebra of partial permutations $\curlyA_{\infty}$ with a non-degenerate topological metric. We further show that the existence of a non-degenerate metric implies that the algebra of operators of Gromov-Witten theory is isomorphic to the Frobenius algebra of partial permutations at infinite $n$. The rest of the section crystallizes the physical ideas in the mathematical correspondence in a manner that may be useful to both communities. 

\subsection{The Gromov-Witten Theory on the Sphere}
\label{GWOnTheSphere}
The backbone of the solution \cite{OP1} of the Gromov-Witten theory is a calculation of the correlators using equivariant localization \cite{Okounkov:2002cz}. The results one needs to describe the correspondence to the Hurwitz (gauge) theory are in the end quite simple. We want to intuitively describe how the simplification comes about, summarizing the  works \cite{OP1,Okounkov:2002cz}.

\subsubsection*{The Reduction to One-point Functions}
Firstly, the relative Gromov-Witten invariants are defined as:
\begin{equation}
\langle \prod_{i=1}^s \tau_{k_i}(\gamma), \xi^1,\dots,\xi^m \rangle^{\circ X}_{g,d} =
\int_{[\overline{M}_{g,s}(X,\xi^i)]^{vir}} \prod_{i=1}^s \psi_i^{k_i} ev_i^\ast (\gamma) \, ,
\label{ModuliSpaceIntegral}
\end{equation}
where the $\tau_{k}(\gamma)$ are  vertex operator insertions associated to the cohomology class $\gamma \in H^*(X)$ of the curve $X$ of genus $g$, the $\xi^i$ are prescribed ramification profiles at $m$ points and we study connected degree $d$ covers of the curve $X$. The correlator is defined as an intersection integral over the moduli space of curves of the appropriately ramified type with an integrand associated to the descendants $\tau_{k_i}(\gamma)$. For further background, see \cite{Witten:1989ig}.

\begin{figure}[ht]
	\centering
	\includegraphics[width=8cm]{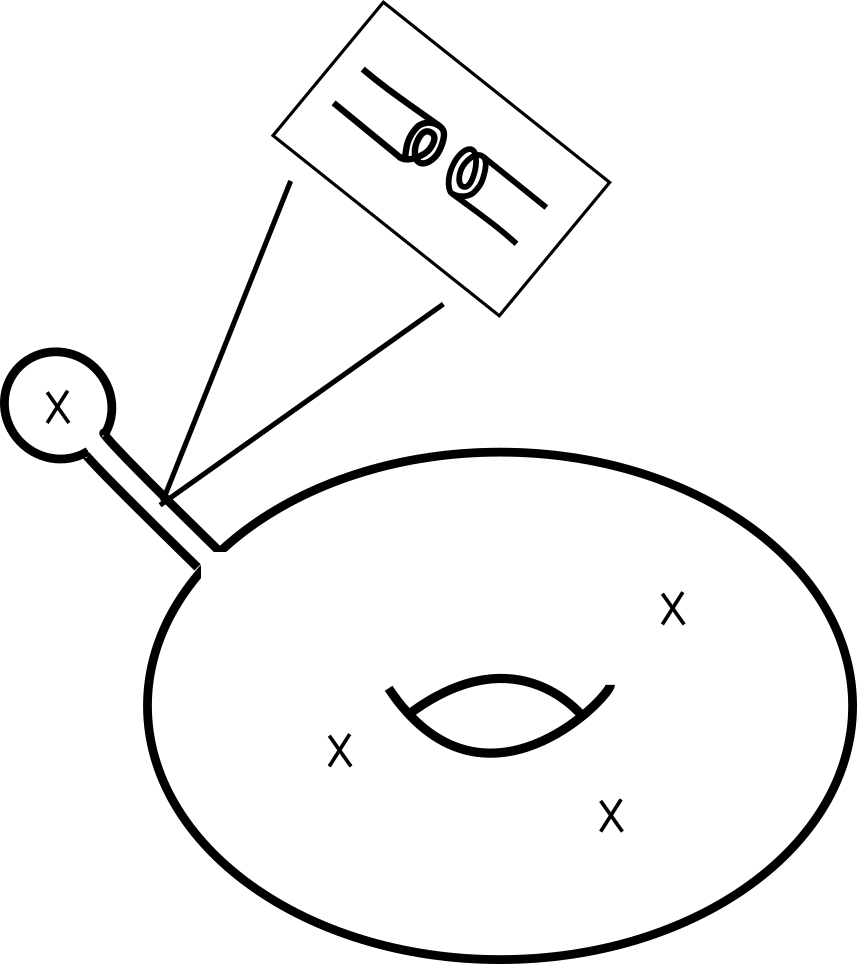}
	\caption{Insertions can be isolated on a two-sphere and glued to the rest of the surface in a manner prescribed by a ramification profile.}
 \label{DegenerationAndGluing}
\end{figure}

We will restrict our attention to the stationary sector of Gromov-Witten theory, 
i.e. to the operator insertions associated to the volume form $\omega$ of the curve $X$. For the disconnected relative correlators indicated with a solid bullet, we have a degeneration formula \cite{GraberPandharipande,Li,Bryan:2004iq}:
\begin{align}
\langle \prod_{i=1}^s \tau_{k_i}(\omega), \xi^1,\dots,\xi^m \rangle^{\bullet X}_{g,d} =
\sum_{|\mu^i|=d} H^X_d(\mu^i,\xi^j) \prod_{i=1}^s z(\mu^i) \langle \mu^i, \tau_{k_i} (\omega) \rangle^{\bullet \PP^1} \, , \label{DisconnectedCorrelators}
\end{align}
where we introduced the order $z$ of the centralizer of a permutation:
\begin{equation}
z(\mu) = \prod_i m_i(\mu) ! i^{m_i(\mu)} \, ,
\end{equation}
and $\langle \mu^i, \tau_{k_i} (\omega) \rangle^{\bullet \PP^1}$ denotes an integral in the stationary sector of $\mathbb{\PP}^1$ relative to $0$.
The degeneration formula implies that the Gromov-Witten calculation can be done by isolating each vertex operator on a two-sphere, and gluing the two-sphere to the rest of the surface in every possible manner, i.e. with any possible partition profile $\mu^i$. See Figure \ref{DegenerationAndGluing}.
One is thus led to the (disconnected, fixed degree) substitution rule that translates operator insertions into ramification profiles represented by permutations \cite{OP1}:
\begin{equation}
\tau_k(\omega) = \sum_{|\mu|=d} z(\mu) \langle \mu, \tau_k(\omega) \rangle^{\bullet \mathbb{P}^1} \cdot (\mu) \, 
. \label{SubstitutionRule}
\end{equation}
This rule is at the core of the correspondence. Furthermore, the disconnected one-point function on the right-hand side of this substitution rule can be rewritten in terms  of connected one-point functions \cite{OP1}:
\begin{equation}
\langle \mu,\tau_k(\omega) \rangle^{\bullet \PP^1} = \sum_{i=0}^{m_1(\mu)} \frac{1}{i!} \langle \mu-1^i,\tau_k(\omega) \rangle^{\circ \PP^1} \, .
\end{equation}
Indeed, if the partition $\mu$ contains $i$ trivial entries, then we can split those entries off, count them for a factor of $1/i!$ \cite{OP1} and consider the rest of the partition $\mu$, i.e. $\mu-1^i$ with the insertion.  
This combinatorics is automatically taken care of by the definition of the generalized Hurwitz numbers. 
To determine the full relative Gromov-Witten theory correlators in (\ref{DisconnectedCorrelators}) one is left with determining generic Hurwitz numbers and connected one-point functions on the sphere. 
 
In summary, we make tubes with $\mathbb{P}^1$'s at the end, for each  vertex operator. Each tube is glued in a particular manner $\mu$ to the rest of the surface -- we sum over all possible manners of gluing. The non-trivial part of the gluing can be of smaller degree than the total degree of the initial surface. This extra possibility will in the end be responsible for the completion of the cycle corresponding to the vertex operator $\tau_k(\omega)$.
To compute amplitudes on a generic genus $g$ Riemann surface with several closed string insertions, we only need the handle operator as well as the one-point functions on $\mathbb{P}^1$ with ramifications. 

\subsection{The One-point Functions and the Completed Cycles}
\label{OnePointAndCompletedCycles}
The one-point functions are computed in an operator formalism in \cite{OP1} based on the equivariant localization of the theory \cite{Okounkov:2002cz}. The calculation takes place in a NS sector fermionic Fock space of a chiral complex fermion. 
The summary of the result is as follows. If we define one-point functions by:
\begin{align}
\frac{\rho_{k+1,\mu}}{k!} &= z(\mu) \langle \mu, \tau_k(\omega) \rangle^{\circ,\PP^1} \, ,
\end{align}
then they are found to be \cite{OP1}
\begin{align}
\rho_{k,\mu} &= (k-1)!  \frac{\prod \mu_i}{|\mu|!}  [z^{k+1-|\mu|-l(\mu)}] S(z)^{|\mu|-1} \prod S(\mu_i z)  \, ,
\end{align}
where
\begin{equation}
S(z) =  \frac{\sinh(\frac{z}{2})}{\frac{z}{2}} \, 
\end{equation}
and $[z^{n}]$ denotes the coefficient of the term $z^{n}$ in an expansion of the expression in powers of $z$. 
The substitution rule (\ref{SubstitutionRule}) then translates  vertex operators into the completed cycles $\overline{(k)}$:
\begin{equation}
\tau_k(\omega) \leftrightarrow 
\frac{1}{k!} \overline{(k+1)}
\end{equation}
where the completed cycle equals:
\begin{equation}
\overline{(k)}= \sum_{\mu} \rho_{\mu,k} (\mu) \, .
\label{HolographicMapObservables}
\end{equation}
This formula agrees with the results we obtained in subsection \ref{CompletedCyclesInHurwitz} for the completed cycles.\footnote{The calculation of the one-point functions proceeds through the equivariantization of the Gromov-Witten theory where there are contributions from nodal curves that correspond to the addition of simple branch points \cite{Okounkov:2002cz}. These intermediate localization results and  exponentials of simple branch points cancel in the non-equivariant limit. It would be good to clarify the physical interpretation of the exponential of the transposition operator that codes these simple branch points as this operator is a natural deformation on the moduli space of string theories, functioning as an elementary string interaction vertex. See e.g. \cite{Dijkgraaf:1997vv,Eberhardt:2021vsx} and references therein for background in matrix theory or $AdS_3/CFT_2$.}

\subsection{The Gauge/String Correspondence}
\label{GSCorrespondence}
Once we established the map of observables (\ref{HolographicMapObservables}), the gauge theory/string theory correspondence follows \cite{OP1}:
\begin{equation}
\prod_{i} \frac{1}{k_i !}\, H^X_d \Big( \overline{(k_1+1)},\dots, \overline{(k_s+1)} \Big)=\langle \prod_i \tau_{k_i} (\omega)
\rangle_d  \, ,
\label{HolographicMapCorrelators}
\end{equation}
at any fixed degree $d$ for the covering maps. We stress that the operator map (\ref{HolographicMapObservables}) is valid at all degrees while the correlators are 
evaluated at fixed degree.

On the left-hand-side, as we explained in section \ref{Hurwitz}, we have a pure gauge theory amplitude of a symmetric group gauge theory in which we identified multiplicative observables based on a power sum basis of shift symmetric functions. On the right-hand side, we have a string theory integral over the moduli space of Riemann surfaces of particular operators -- see equation (\ref{ModuliSpaceIntegral}).

\subsubsection*{One Step Beyond}
We can proceed one step further in our interpretation of the duality. We note that the degree $d$ of the cover of the sphere $\mathbb{P}^1$ is equal to the world sheet instanton number. We can weigh each instanton of degree $d$ with a factor $q^d$, where we imagine that $q$ is proportional to the exponential of (minus) the instanton action (proportional to the complexified area of the target sphere). The correlation functions of  string theory are given by such weighted sums. Thus, it is natural to define correlators on the extended algebra $\curlyA_{\infty}$ of partial permutations by taking weighted sums of correlation functions of all orders:
\begin{equation}
    \langle \cdot \rangle := \sum_{d=0}^\infty q^d \langle \cdot \rangle_d \, .
    \label{NonDegenerateBilinearForm}
\end{equation}
This definition gives rise to a non-degenerate two-point function on the algebra. To prove the non-degeneracy, we show that for any given element $\sum_r \kappa_r A_r$ of $\curlyA_{\infty}$, we can find an element $A_s$ for which the two-point function
is non-zero. Denote by $\bar{s}$ the minimal partial permutation that completes to $\sigma$. The only terms in the sum  $\sum_r \kappa_r A_r$ that contribute to the two-point function $\langle \left( \sum_r \kappa_r A_r \right) A_s \rangle$ are those for which the partial permutation $r$ is of the form $\bar{s}\cup 1^l$. Let $l_{min}$ denote the minimal value of $l$ for which the coefficient $\kappa_{r\cup 1^l}$ is non-zero. We write
\begin{equation}
    \langle \left( \sum_r \kappa_r A_r \right) A_{\bar{s}\cup 1^{l_{min}}} \rangle = \kappa_{\bar{s}\cup 1^{l_{min}}} q^{\lvert d_{\bar{s}} \rvert + l_{min}}\cdot (1+\mathcal{O}(q))\, .
\end{equation}
This two-point function is non-zero for a generic (non-zero) value of $q$. The right-hand-side defines an analytic function which can have up to a countably infinite number of zeros. Appendix \ref{AnExplicitProof} contains another proof which rules out this latter possibility: it demonstrates explicitly that the bilinear form is non-degenerate for any non-zero value of $q$.

The bilinear form (\ref{NonDegenerateBilinearForm}) thus defines a Frobenius structure 
on the extended algebra $\curlyA_{\infty}$. The resulting Frobenius algebra defines a 
new topological field theory that captures all Hurwitz theories at finite $n$ exactly 
and provides a more natural setting for the gauge theory/string theory correspondence. 
In particular, the algebra of vertex operators of this string theory is isomorphic to 
the algebra of orbits of partial permutations $\curlyA_{\infty}$ and the correlators 
summed over instanton degree match.
\subsubsection*{Lessons Learnt}
We want to stress a few lessons we learn from this rather uniquely exact and explicit correspondence. Firstly, if on the string theory side, we integrate over open moduli spaces, the right-hand side of equation (\ref{HolographicMapCorrelators}) will reduce to Hurwitz numbers of simple cycles. This is the standard intuition for bulk string vertex operators corresponding to single trace or single cycle operators in the gauge theory. However, we see that if we compactify the  moduli space, we need to fine tune this intuitive picture to define generalizations of the single cycle operators which are the completed cycles. Completion originates in compactification which in turn is necessary because of the degeneration of surfaces on the boundary of moduli space. 
Secondly, the map of the algebra of observables (\ref{HolographicMapObservables}) is $n$-independent but the evaluation of the correlators gives finer results at given finite $n$.  
Thirdly, the instanton sum on the string theory side provides strong motivation to sum over the gauge theories with gauge group $S_n$ for all values of $n$, moreover providing the latter theory with a non-degenerate topological metric.\footnote{The grand canonical approach to the $S_n$ dual of a string theory also appears natural in the context of the evaluation of three-dimensional gravitational partitions functions and their symmetric orbifold duals \cite{Eberhardt:2020bgq}.}

To what extent can this gauge/string correspondence be extended to a general orbifolded topological quantum field theory ? We observe that the graded Hurwitz theory is dual to the genus $0$ restriction of Gromov-Witten theory on $\mathbb{P}^1$. 
Moreover, we argued that the graded theory codes a quotient of the second quantized algebra. Thus, there is a subsector of the general theory that is subject to the duality. 
Also, we note that there is a generalization of a free field that governs the infinite $n$ algebra of observables in the general case\footnote{See formula (4.9) in \cite{Ashok:2023kkd}.} providing an infinite $n$ starting point for a large $n$ expansion.\footnote{We develop a large $n$ point of view on the gauge/string duality for the Hurwitz case in Appendix \ref{LargeNHurwitz}.} The latter is still governed by a clear construction of a covering surface. These initial remarks spur hope that there is again a duality map for the operator algebra and  correlation functions in the general case.

\subsubsection*{Large $n$ Counting}
Finally, let us make a few remarks on the nature of the amplitudes that are computed in the topological theory and their large $n$ counting. The Riemann-Hurwitz formula or the dimension constraint on the intersection number on moduli space gives rise to the constraint
\begin{equation}
2g-2 +d (2-2 g(X)) = \sum_{i=1}^s k_i \, 
\end{equation}
for the Gromov-Witten amplitude on a target curve $X$ of genus $g(X)$. In this paper, we have concentrated on $X=\mathbb{P}^1$. 
The first remark is that for a fixed set of  operator insertions, at each instanton number there is maximally one genus $g$ that contributes to the amplitude. The string coupling dependence for a given amplitude will thus be the string coupling to the power:
\begin{equation}
2g-2+s = \sum_{i=1}^s (k_i+1) - 2d \, .
\end{equation}
This is therefore also the expected power of $1/\sqrt{n}$ of the amplitude. Indeed, normalized amplitudes behave in this manner as argued in detail in 
\cite{Pakman:2009zz}. 
For single cycle operators of length $n_i=k_i+1$, the normalization factors give rise to a leading factor of  $n^{-\frac{ \sum_i n_i}{2}}$ and there is a choice of  $n-d$ colours not participating in the interaction which at large $n$ gives a factor of choice of $n^{d}$. 
This is precisely as expected for a string coupling $g_s=n^{-1/2}$ at Euler number $\chi=2-2g$ and with $s$ external states.\footnote{In this reasoning we are to think of multi-cycle operators as a product of single cycle operators with a relative factor of order one.} 
We conclude that in the topological theory, the large $n$ behaviour of a given amplitude is set by the instanton number and vice versa. 

\section{Conclusions}
\label{Conclusions}
In this paper, we discussed two-dimensional topological quantum field theories with $S_n$ gauge symmetry. We stressed that mathematical results \cite{OP1} on the pure gauge theory allow for an interpretation as an exact gauge/string duality. Through degeneration \cite{Li} correlators can be reduced to relative one-point functions \cite{OP1} and localization \cite{Okounkov:2002cz} then allows for their exact evaluation. In the process, the string vertex operators can be matched to a multiplicative basis of completed cycles \cite{OP1}. We stressed that these completed cycle operators live in the algebra of partial permutations \cite{IvanovKerov} defined at infinite $n$. 
Furthermore, we showed that this algebra can be endowed with a non-degenerate metric by summing the correlators over all possible degrees of the covering map, which suggests that the proper setting for this gauge/string duality is a grand canonical field theory. We generalized the construction of the extended operator algebra to all symmetric orbifolds of topological quantum field theories in two dimensions. We were able to separate the $n$-dependence of the structure constants from their dependence on the structure constants of the seed Frobenius algebra. There is justified hope that a large $n$ limit of these theories will also lay bare a string theory dual. 

\begin{figure}[ht]
	\centering
	\includegraphics[width=5cm]{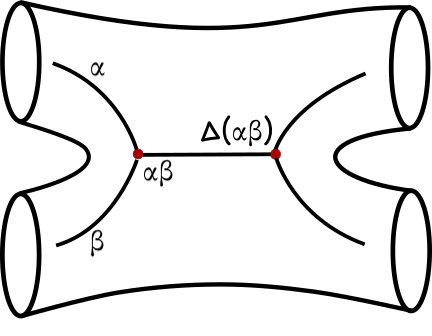}
	\caption{A joining and splitting interaction decorated with seed Frobenius algebra structure constants.}
 \label{ThreePointExample1}
\end{figure}

\begin{figure}[ht]
	\centering
	\includegraphics[width=5cm]{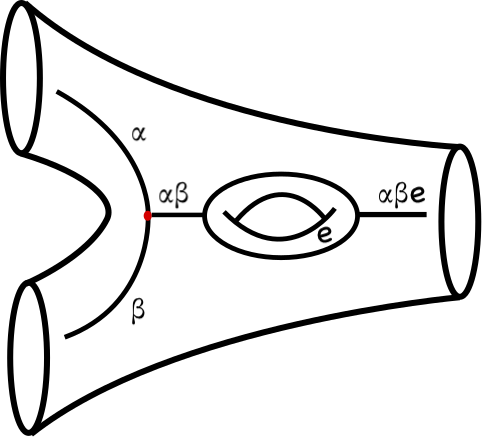}
	\caption{A one loop joining with a seed handle operator.}
 \label{ThreePointExample2}
\end{figure}

The results of this paper connect to the literature in many manners and allow for interesting extensions. We discuss a few of these avenues. The topological quantum field theory technology has already allowed for the calculation of a very large class of extremal correlators \cite{Li:2020zwo,Ashok:2023mow,Ashok:2023kkd}
that were out of reach of Feynman diagram techniques \cite{Pakman:2009ab,Pakman:2009zz}. Our abstract treatment of the topological symmetric orbifold theories should lead to the formulation of new Feynman rules that further simplify the calculation of extremal correlators. We sketch such Feynman diagrams on the covering surface with a skeleton of seed Frobenius interactions in Figures \ref{ThreePointExample1} and \ref{ThreePointExample2}.

\noindent These illustrate the calculation of the operators in terms four and five on the right-hand side of the following equation 
\begin{equation}
\begin{split}
    O_{3}(\alpha) \cdot O_{3}(\beta) = 2O_{3,3}(\alpha,\beta)& +5O_{5}(\alpha\beta)  +8O_{2,2}(\Delta_{*}^{(2)}(\alpha \beta))+O_{3}(\alpha \beta e)\\ &+3(n-3) O_{3}(\Delta_*^{(2)}(\alpha \beta))+\frac{n(n-1)(n-2)}{3} O_{e}(\Delta^{(3)}_*(\alpha \beta))\, . \label{Example33}
\end{split}
\end{equation}
For now, these diagrams are schematic, but we believe that they can be turned into an efficient tool to paint the necessary seed Frobenius algebra calculations onto the covering surface of the topological orbifold.

A second motivating connection between our results and the literature is as follows. In
\cite{Eberhardt:2019ywk,Eberhardt:2021vsx} it has been convincingly argued that the dual of the string scale superstring in $AdS_3\times S^3\times M$ is given by the symmetric orbifold theory $\text{Sym}_n (M)$. The topological subsector of this duality may be   amenable to an (even more) direct proof of the duality. 
In particular, it has been shown that a subclass of the operators of this theory, i.e. the chiral ring of the physical theory \cite{Lerche:1989uy}, can be matched to conjugacy
classes of the symmetric group \cite{Lunin:2000yv,Lunin:2001pw} allowing for the proof 
of conjectured correlators and the efficient calculation of many more \cite{Li:2020zwo,Ashok:2023mow,Ashok:2023kkd}. 
One would like to make this connection even more fluent, firstly, by rigorously proving the localization onto holomorphic covering maps using supersymmetry and secondly, by identifying the insertion of vertex operators at the boundary with the gluing of boundary points, as in the theory of world sheet instanton contributions to topological sigma-models \cite{Witten:1989ig}. 
Given the close connection between the descendant operators $\tau_k(\omega)$ and world sheet instantons \cite{Witten:1989ig}, this seems like a direct path to identifying our topological symmetric orbifold as a subsector of the $AdS_3/CFT_2$ correspondence.

Finally, we remark that there is a triangle of equivalences between three theories: the quantum cohomology of the Hilbert scheme of points on $\mathbb{C}^2$, the Gromov-Witten theory on $\mathbb{P}^1 \times \mathbb{C}^2$ and the Donaldson-Thomas theory on $\mathbb{P}^1 \times \mathbb{C}^2$. See e.g. \cite{OPHilbert}.  Moreover, all vertices of the triangle allow for an equivariant counterpart.
We mainly comment on the relation between the first two legs of the triangle which are more directly related to the topic of this paper. There is a direct relation between relative Gromov-Witten invariants on $\mathbb{P}^1 \times \mathbb{C}^2$ and the quantum  intersection numbers for cohomology classes on the Hilbert scheme of points on the affine plane. These are described in terms of ordinary (not shifted) symmetric function theory. 
It would be interesting to clarify whether completed cycles can also play a role in this context, e.g. in the absolute Gromov-Witten theory. Interesting explorations in this research direction were already recorded in \cite{Lerche:2023wkj}. This is one of the many open ends of our work that deserves to be explored further. 

\appendix
\section{Alternative Completed Cycles}
\label{AlternativeCompletedCycles}
In this Appendix, we list the completed cycles up to length six that arise from identifying the trivial completed cycle as the trivial cycle (i.e. changing the constant shift in the definition of the power sum polynomials). The expressions are simpler:
\begin{align}
    (\bar{1}) &=(1)\\
    (\bar{2}) &= (2) \nonumber \\
    (\bar{3}) &= (3)+(1,1)+\frac{1}{12}\cdot (1)\nonumber\\
    (\bar{4}) &= (4)+2\cdot (2,1)+\frac{5}{4}\cdot (2) \nonumber\\
    (\bar{5}) &= (5)+3\cdot (3,1)+4\cdot (2,2)+\frac{11}{2}\cdot (3)+4\cdot (1^3) +\frac{3}{2}\cdot (1,1)+\frac{1}{80}\cdot (1) \nonumber\\
    (\bar{6}) &= (6) + 4\cdot (4,1) + 6 \cdot (3,2) +\frac{95}{6} (4) + 10\cdot (2,1,1) +\frac{35}{3}\cdot (2,1)+ \frac{91}{48}\cdot (2) \, . \nonumber
\end{align}

\section{Conjugacy Class Sums and the Large 
\texorpdfstring{$n$}{n}  Limit}
\label{InfiniteN}
\label{LargeNHurwitz}
In this Appendix, we study the large $n$ limit of Hurwitz theory and discuss the extent to which there is an analogy with the 't Hooft limit of gauge theories. Firstly, we consider the zeroth order approximation to the large $n$ behaviour of the Hurwitz theory. Secondly, we explain how the correlators of this theory are encapsulated by complicated string correlators. We conclude with a comparison between the gauge/string duality discussed in the bulk of this paper and the conventional point of view on gauge/string dualities.

The product (\ref{Product}) and correlation functions (\ref{CorrelationFunctions}) we have defined for the Hurwitz theory diverge in the large $n$ limit. This is merely an artefact of the normalization of our operators and we can define operators $\Phi_{\eta}$ that satisfy:
\begin{equation}
   \langle \Phi_{\eta}\Phi_{\sigma} \rangle=\delta_{[\eta],[\sigma]}\, .
\end{equation}
The comparison with the two-point function (\ref{MetricOnTheSphere}) leads to 
the definition 
\begin{equation}
    \Phi_{\eta}=\frac{C_{\eta}}{\sqrt{\lvert C_{\eta} \rvert}}\, .
    \label{Renormalization}
\end{equation}
These operators form a basis of the conjugacy class algebra $\curlyC_n$. At 
large $n$, their products and correlators remain finite, and can be expanded in $1/\sqrt{n}$ \cite{Keller}. The powers of $n$ associated to each term in their expansion
are given by $n^{-n_3/2}$, where $n_3$ denotes the triple overlap of three permutations 
in each of the three conjugacy classes \cite{Keller}. The highest order term in the expansion can be computed explicitly for single cycles
\begin{equation}     
    \Phi_{i}\Phi_{j}=
    \begin{cases}
    \Phi_{[i,j]} +\mathcal{O}(1/\sqrt{n}) &\quad \text{if $i \neq j$}\\
    \sqrt{2} \Phi_{[i,i]} + \Phi_e +\mathcal{O}(1/\sqrt{n}) &\quad \text{if $i=j$}
    \end{cases}
    \, .
\end{equation}
This shows that the correlation functions of the large $n$ theory can be computed using Wick contractions, as operators must match pairwise in order for the correlators not to vanish. Moreover, we note that the leading large $n$ behaviour shows that it is crucial in our context to allow for both connected and disconnected string diagrams. Indeed, the leading term in an operator product corresponds to a disconnected string diagram. 

Generalizing this result to the product of a single-cycle operator with a multi-cycle operator, we find
\begin{equation}
    \Phi_{i} \Phi_{i^{k}j^{l_1}\dots j^{l_s}}=\sqrt{k+1}\Phi_{i^{k+1}j^{l_1}\dots j^{l_s}}+\sqrt{k}\Phi_{i^{k-1}j^{l_1}\dots j^{l_s}}
\end{equation}
A consequence is that the large $n$ limit of the operator algebra of Hurwitz theory admits a faithful representation in a bosonic Fock space\footnote{It would be interesting to understand the link between the Fock space we have defined here and the fermionic Fock space of Hurwitz theory \cite{OP1}.} 
\begin{equation}
    \Phi_i \mapsto a_i+a_{i}^{\dag} \, .
    \label{FockSpaceRep}
\end{equation}
One can think of the right-hand side of \ref{FockSpaceRep} as describing the degrees of freedom of a real, two-dimensional free bosonic field at a given point. 

Continuing to think in terms of these traditional single cycle operators, we can provide a different perspective on the operator map (\ref{HolographicMapCorrelators}) that matches correlators. Completed cycles are a basis of the partial 
permutation algebra $\curlyA_{\infty}$. Therefore, single cycles can be written as 
linear combinations of completed cycles and matched to complicated linear combinations 
of products of vertex operators in the string theory:
\begin{equation}
    A_{k}=\sum_{\substack{l\leq k\\\mu\vdash l}} \zeta_{\mu} \prod \tau_{\mu_{i}}(\omega) \, .
    \label{ReversedHolographicMap}
\end{equation}
The coefficients $\zeta_{\mu}$ in this expression are obtained by inverting the change-of-basis matrix between shifted power sums and normalized characters. 
A consequence of the map (\ref{ReversedHolographicMap}) is that correlators of products of single cycles can be matched to that of products of complicated string operators at any fixed order $n$. The large $n$ limit may then be taken and we find that all Hurwitz correlators are encoded in the string theory. However, this map is not bijective: the string theory contains far more operators than the Hurwitz theory (at fixed $n$) does. Establishing a one-to-one duality requires extending the Hurwitz algebra to that of orbits of partial permutations, as we have explained in the bulk of the paper. 

We conclude that the Hurwitz theory is entirely encoded in the fixed-degree correlators of a string theory. However, the latter contains information on all degrees at once. Therefore, in order to get a gauge/string duality, it is imperative to take a grand canonical approach to the field theory side.

\section{A  Proof of Non-Degeneracy}
\label{AnExplicitProof}
In this Appendix, we provide a slightly stronger proof of the non-degeneracy of the two-point function (\ref{NonDegenerateBilinearForm}). The strategy is to show explicitly that there is no non-trivial element $\sum_{r} \kappa_r A_r$ of the algebra $\curlyA_{\infty}$ whose two point function with every basis vector $A_s$ vanishes. Denote by $\bar{s}$ a minimal partial permutation. We compute
\begin{align}
    \Big \langle \left(\sum_{r} \kappa_r A_r \right) A_{\bar{s}\cup 1^l} \Big \rangle :=\, & \sum_{r,d} q^d \kappa_r \langle A_r A_{s\cup 1^l} \rangle_d 
    \label{NonDegeneracyOfTheMetric}\\
     =\, & \sum_{r,d} q^d \kappa_r \binom{m_1(\rho)}{m_1(r)} \binom{d-\lvert d_{\bar{s}} \rvert}{l} \lvert [\rho;d] \rvert \delta_{[\rho],[\sigma]}
     \nonumber \\
    =\, & \sum_{\substack{k=0  \rvert \\d\geq \lvert d_{\bar{s}}\rvert}} ^{k=d-\lvert d_{\bar{s}}}\kappa_{\bar{s}\cup 1^k} q^d \binom{d-\lvert d_{\bar{s}} \rvert}{k} \binom{d-\lvert d_{\bar{s}} \rvert}{l} \lvert [\sigma;d] \rvert 
    \nonumber \\
    =\, & \frac{1}{\prod_{i\geq 2} m_i{s}! i^{m_i(s)}} \sum_{k\geq 0} \kappa_{\bar{s}\cup 1^k} \left(\sum_{d\geq \lvert d_{\bar{s}}\rvert} q^d \binom{d-\lvert d_{\bar{s}} \rvert}{k} \binom{d-\lvert d_{\bar{s}} \rvert}{l} \frac{d!}{(d-\lvert d_s \rvert)!} \right)\, . \nonumber
\end{align}
The terms inside the parentheses can be thought of as vectors $v_k$ indexed by the non-negative integer $k$. To clarify the discussion, we introduce a bound $D$ that we take to infinity at the end of the computation. The number of vectors in the sum equals the dimension of the vector space they live in. The vector space  has dimension $D-\lvert \bar{d}_s \rvert$. 

To prove that the two-point function (\ref{NonDegeneracyOfTheMetric}) vanishes only 
if all the coefficients $\kappa$ do, we will show that the vectors $v_k$ are linearly independent. To that end, we compute the determinant of the matrix\footnote{Note that $\theta_d$ kills the partial permutations whose support has a cardinal bigger than $d$.
This fact is implemented in our computations in the properties of the binomial, as $\binom{n}{k}$ vanishes when $k>n$. This is why we are allowed to restrict the sum 
(\ref{MatrixDeterminant}) to the numbers $d$ that satisfy $d\geq \lvert d_{\bar{s}}\rvert+k$.} 
\begin{equation}
    \left(\sum_{d\geq \lvert d_{\bar{s}}\rvert+k} q^d \binom{d-\lvert d_{\bar{s}} \rvert}{k} \binom{d-\lvert d_{\bar{s}} \rvert}{l} \frac{d!}{(d-\lvert d_s \rvert)!} \right)_{k,l}\, .
    \label{MatrixDeterminant}
\end{equation}
Our strategy is to perform the following operations on the lines $L_k$ of the matrix, from the bottom up:
\begin{equation}
    L_k - \frac{\binom{d-\lvert d_{\bar{s}}\rvert}{k}}{\binom{d-\lvert d_{\bar{s}} \rvert}{p}} L_{p\geq n}\, .
\end{equation}
That is, start by substracting $L_n$ from
$L_{1},\dots, L_{n-1}$ with the appropriate coefficients given above. Then, in the newly obtained matrix, substract $L_{n-1}$ from lines $L_{1},\dots, L_{n-2}$, and so on. The determinant of the matrix (\ref{MatrixDeterminant}) reduces to the determinant of the matrix 
\begin{equation}
    \left(q^d \binom{k}{l} \frac{(\lvert d_{\bar{s}}\rvert+k)!}{k!} \right)_{k,l}\, .
\end{equation}
This is a lower triangular matrix whose determinant is given by the product of the diagonal elements. For non-zero $q$, no such elements vanish and we thus find that the determinant of the matrix (\ref{MatrixDeterminant}) is non-zero. In consequence, the two-point functions \ref{NonDegeneracyOfTheMetric} vanish identically only for the trivial 
element $0$. They are non-degenerate.

\end{document}